\documentclass[twocolumn,epjc3]{svjour3} 

\RequirePackage{graphicx}
\RequirePackage{mathptmx}      
\RequirePackage{flushend}
\RequirePackage[numbers,sort&compress]{natbib}
\RequirePackage[colorlinks,citecolor=blue,urlcolor=blue,linkcolor=blue]{hyperref}

\usepackage{amsmath}
\usepackage{amssymb}

\smartqed  

\usepackage{array}
\usepackage[utf8]{inputenc}
\usepackage{latexsym}
\usepackage{color}
\usepackage{hyperref}
\usepackage{graphics}
\usepackage{subcaption}
\usepackage{boldline}
\usepackage{multirow}
\usepackage{rotating}

\usepackage{enumitem}
\usepackage{url}

\newcommand{\be}{\begin{equation}}
\newcommand{\ee}{\end{equation}}
\newcommand{\bea}{\begin{eqnarray}}
\newcommand{\eea}{\end{eqnarray}}

\newcommand{\lsim}{\raise0.3ex\hbox{$\;<$\kern-0.75em\raise-1.1ex
\hbox{$\sim\;$}}}
\newcommand{\gsim}{\raise0.3ex\hbox{$\;>$\kern-0.75em\raise-1.1ex
\hbox{$\sim\;$}}}

\let\oldhref\href
\renewcommand{\href}[2]{\oldhref{#1}{\hbox{#2}}}

\journalname{Eur. Phys. J. C}

\begin{document}

\title{Decoherence in neutrino propagation through matter, and bounds from IceCube/DeepCore}

\author{P. Coloma\thanksref{e1,addr1}\and J. Lopez-Pavon\thanksref{e2,addr2}\and I. Martinez-Soler\thanksref{e3,addr3},
H. Nunokawa\thanksref{e4,addr4}
}

\thankstext[]{t1}{CERN-TH-2018-041; IFT-UAM/CSIC-18-022;\\ FERMILAB-PUB-18-067-T}
\thankstext{e1}{pcoloma@fnal.gov}
\thankstext{e2}{jacobo.lopez.pavon@cern.ch}
\thankstext{e3}{ivanj.m@csic.es}
\thankstext{e4}{nunokawa@puc-rio.br}

\institute{Theory Department, Fermi National Accelerator Laboratory, P.O. Box 500, Batavia, IL 60510, USA \label{addr1} 
          \and Theoretical Physics Department, CERN, 1211 Geneva 23, Switzerland\label{addr2}
          \and Instituto de F\'{\i}sica Te\'orica UAM-CSIC, Calle Nicolas Cabrera 13-15, Cantoblanco E-28049 Madrid, Spain\label{addr3}
          \and Departamento de F\'isica, Pontif\'icia Universidade Cat\'olica do Rio de Janeiro, C. P. 38097, 22451-
900, Rio de Janeiro, Brazil\label{addr4}          
}

\date{Received: date/ Accepted:date}

\maketitle

\abstract{
We revisit neutrino oscillations in matter considering the open quantum
system framework, which allows to introduce possible decoherence effects
generated by 
New Physics in a phenomenological manner. We assume that the decoherence parameters $\gamma_{ij}$ may depend on the neutrino energy, as
$\gamma_{ij}=\gamma_{ij}^{0}(E/\text{GeV})^n$ $(n = 0,\pm1,\pm2) $.  
The case of 
non-uniform matter is studied in detail and, in particular, we develop a consistent formalism to study the non-adiabatic 
case dividing the matter profile into an arbitrary number of 
layers of constant densities. This formalism is then applied to explore 
the sensitivity of IceCube and DeepCore 
to this type of effects. Our study is the first atmospheric 
neutrino analysis where a consistent treatment of the matter effects in the three-neutrino case 
is performed in presence of decoherence. We show that matter effects are indeed extremely relevant in this context. We find that
IceCube is able to considerably improve over current bounds 
in the solar sector ($\gamma_{21}$) and in the atmospheric sector ($\gamma_{31}$ and $\gamma_{32}$) for $n=0,1,2$ and, in particular, by several 
orders of magnitude (between 3 and 9) for the $n=1,2$ cases. For $n=0$ we 
find $\gamma_{32},\gamma_{31}< 4.0\cdot10^{-24} (1.3\cdot10^{-24})$ GeV and 
$\gamma_{21}<1.3\cdot10^{-24} (4.1\cdot10^{-24})$ GeV at the 95\% CL, for normal (inverted) mass ordering. 
}

\section{Introduction}
\label{sec:intro}
The accurate measurement of the mixing angle $\theta_{13}$ 
by reactor neutrino experiments~\cite{An:2015rpe}, 
with a small uncertainty 
comparable to that for $\theta_{12}$, has initiated a precision era for neutrino 
physics. In the standard three-family framework, the main remaining issues are the possible observation of leptonic CP violation, 
the determination of the ordering of neutrino masses and probing the Dirac or Majorana nature of neutrinos. Some hints currently exist 
in the latest data collected by NOvA and T2K which seem to point to maximal CP violation in the neutrino sector, but the statistical 
significance is still low~\cite{T2K-2017,NOvA-Seminar-FNAL-2017}. Likewise, a global fit to neutrino oscillation data seems to show a mild preference
for a normal mass ordering (see for instance~\cite{nufit,Esteban:2016qun}), which needs to be confirmed as more data become available. 

At the same time, and in view of the precision of present and near future neutrino facilities, it is of key importance to verify if neutrinos have 
unexpected properties caused by New Physics (NP) beyond the standard three-family framework.
In this work we study one of the possible windows to NP, the so-called quantum decoherence in neutrino oscillations, and update the existing bounds by analyzing IceCube and DeepCore data on atmospheric neutrinos.
In particular, we are interested in a kind of decoherence effects in neutrino
oscillations studied, for example, in
\cite{Benatti:2000ph,Lisi:2000zt,Gago:2000qc,Gago:2002na,Morgan:2004vv,Hooper:2004xr,Fogli:2007tx,Farzan:2008zv}
and, more recently, in \cite{Bakhti:2015dca,Gomes:2016ixi,Oliveira:2016asf,Coelho:2017zes,Coelho:2017byq,Carpio:2017nui}.
These decoherence effects differ from the standard decoherence caused by the
separation of wave packets (see e.g.~\cite{Giunti:1997wq, An:2016pvi}) and might arise, instead, from 
quantum gravity effects~\cite{Hawking:1976ra,Ellis:1983jz,Giddings:1988cx}. 
Throughout this work, for brevity, we will refer to such non-standard decoherence simply as
decoherence. 

The authors of ref.~\cite{Lisi:2000zt} derived some of the strongest available constraints on neutrino decoherence in neutrino oscillations up to date, using atmospheric neutrino data from the Super-Kamiokande (SK) experiment~\cite{Fukuda:1998mi,Fukuda:1998tw,Fukuda:1998ub,Fukuda:1998ah}. Moreover, they considered the general case in which the decoherence parameters could 
depend on the neutrino energy via a power law,
$\gamma=\gamma_{0}(E/\text{GeV})^n$, where $n = 0, -1, 2$.
Nevertheless, these limits were obtained within a
simplified two-family framework and without taking into account the matter
effects in the neutrino propagation. Furthermore, only a reduced subset of SK data (taken, in fact, almost 20 years ago now) was 
analyzed~\cite{Fukuda:1998mi,Fukuda:1998tw,Fukuda:1998ub,Fukuda:1998ah}. 

In this work, we show that performing a three-flavour analysis 
which includes the matter effects is essential in order to correctly interpret such constraints.
In particular, it is not obvious  
to which $\gamma_{ij}$ parameter the SK bounds derived in two
families~\cite{Lisi:2000zt} would actually apply. 
We will show that it strongly depends 
on the neutrino mass ordering and on whether the sensitivity 
is dominated by the neutrino or antineutrino channels: 
for neutrinos the decoherence effects at high energies 
are mainly driven by $\gamma_{21}$ ($\gamma_{31}$) for normal (inverted)
ordering, while in the antineutrino channel they are essentially
controlled 
by $\gamma_{32}$ ($\gamma_{21 }$) for normal (inverted) ordering. 
Concerning the solar sector, the authors of ref.~\cite{Gomes:2016ixi} obtained strong constraints on $\gamma_{21}$ from an analysis of KamLAND data~\cite{Eguchi:2002dm, Abe:2008aa}, 
for $n=0,\pm1$.\footnote{It should be mentioned that, in~\cite{Fogli:2007tx}, 
very strong bounds on dissipative effects were derived from solar neutrino data, 
for $n=0,\pm1,\pm2$ and in a two-family approximation. However, such limits do not apply to the case in which only decoherence 
effects are included, as pointed out in~\cite{Oliveira:2014jsa,Gomes:2016ixi}. 
This will be further clarified in sec.~\ref{sec:adiabatic}.}  
Finally, the authors of ref.~\cite{deOliveira:2013dia} derived several bounds on the atmospheric decoherence parameters $ \gamma_{32}$ and $\gamma_{31}$ from 
an analysis of MINOS data~\cite{Adamson:2011fa, Adamson:2011ig, Adamson:2012rm}. 

Non-standard decoherence has been invoked several times in the
literature in order to decrease the tension in the parameter space
among different sets of neutrino oscillation data. For example, in
refs.~\cite{Farzan:2008zv,Bakhti:2015dca} a solution to the LSND
anomaly based on quantum decoherence, compatible with global neutrino
oscillation data, was proposed. More recently,
in~\cite{Coelho:2017zes} it was shown that the $\sim 2\sigma$ tension
between T2K and NOvA on the measurement of the atmospheric mixing
angle $\theta_{23}$ could be alleviated through the inclusion of
decoherence effects in the atmospheric neutrino sector, namely,
$\gamma_{23} = (2.3 \pm 1.1)\cdot 10^{-23}$ GeV. Such value of
$\gamma_{23}$ would be close to the SK bound from
ref.~\cite{Lisi:2000zt}, $\gamma < 3.5 \cdot 10^{-23}$ GeV ($90\%$
CL), but still allowed.  This topic has recently brought the attention
of a part of the community. In fact, several analyses of decoherence
effects on present and future long-baseline neutrino oscillation
experiments have been recently performed (albeit at the probability
level only), see
e.g. refs.~\cite{Oliveira:2016asf,Coelho:2017byq,Carpio:2017nui}. In
this work we will show that the reference value for $\gamma_{23}$
considered in~\cite{Coelho:2017zes} is indeed already excluded by
IceCube data (we note however that, according to the latest results
reported by NOvA, the significance of the tension has been reduced to
less than 1$\sigma$~\cite{NOvA-Seminar-FNAL-2017}).  Moreover, we find
that IceCube and DeepCore data are able to improve significantly over
most of the constraints in past literature, both for solar and
atmospheric decoherence parameters, in some cases by several orders of
magnitude.

The paper is structured as follows. In sec.~\ref{sec:formalism} we
present the formalism and discuss the effects of decoherence on the
oscillation probabilities. We first review the case of constant matter
density profile, and then proceed to discuss the case of non-uniform
matter. In particular we show that, within the adiabatic
approximation, no significant bounds on the decoherence parameters can
be extracted from solar neutrino data when the neutrino energy is
assumed to be conserved.  We then proceed to develop a formalism which
permits a consistent treatment of the decoherence effects on neutrino
propagation in non-uniform matter when the adiabaticity condition is
not fulfilled, as is the case of atmospheric neutrino experiments. In
sec.~\ref{sec:plots-probabilities} we apply this formalism to the
computation of the relevant oscillation probabilities in the
atmospheric neutrino case, discussing the main features arising in
presence of decoherence.  Section~\ref{sec:icecube} summarizes the
main features of the IceCube and DeepCore experiments, the data sets
considered in our analysis, and the details of our numerical
simulations. Our results are then presented and discussed in
sec.~\ref{sec:results}. Finally, we summarize and draw our conclusions
in sec.~\ref{sec:conclusions}. \ref{app}
and~\ref{sec:d5results} discuss technical details regarding some of
the approximations used in our numerical calculations.

\section{Quantum decoherence: Density matrix formalism}
\label{sec:formalism}

The evolution of the density matrix $\rho$ in the neutrino system can
be described as

\be
\frac{d\rho}{dt}=-i\left[H,\rho \right]-\mathcal{D}\left[\rho\right],
\label{eq:1}
\ee
where $H$ is the Hamiltonian of the neutrino system and the second
term $\mathcal{D}\left[\rho\right]$ parameterizes the decoherence
effects. In vacuum, the diagonal elements of the Hamiltonian are given
by $h_i=m_i^2/(2E)$, where $m_i\ (i=1,2,3)$ are the masses of the
three neutrinos and $E$ is the neutrino energy. Here $\rho$ is defined
in the flavour basis, with matrix elements
$\rho_{\alpha\beta}$. Throughout this work, we will use Greek indices
for flavor ($\alpha,\beta=e,\mu,\tau$), and Latin indices for mass
eigenstates ($i,j = 1,2,3$).

A notable simplification of eq.~\eqref{eq:1} can be achieved via the following set of assumptions. First, assuming complete positivity, the decoherence
term $\mathcal{D}\left[\rho\right]$ can be written in the so-called
Lindblad form~\cite{Lindblad:1975ef,Banks:1983by}
\be
\mathcal{D}\left[\rho\right]=\sum_m\left[\left\{\rho,D_m D_m^\dagger\right\}-2D_m\rho D_m^\dagger\right],
\label{eq:Dm}
\ee
where $D_m$ is a general complex matrix. Second, avoiding unitarity
violation, which is equivalent to imposing the condition $d\,\text{Tr}[\rho]/dt=0$, requires $D_m$ to be Hermitian. Moreover,
$D_m=D_m^\dagger$ implies that the entropy $S=\text{Tr}[\rho\,\ln\rho]$
increases with time.  Finally, a key assumption is the average energy
conservation of the neutrino system, which is satisfied when 

\be
[H,D_m]=0.
\label{eq:[H,D]}
\ee

In presence of matter effects, the Hamiltonian is diagonalized by the unitary mixing matrix\footnote{Note that we consider the standard definition for 
the relation between the mass and flavour eigenstates used in neutrino oscillations: for field operators, $\nu_\alpha =
 \sum_i U_{\alpha i}\nu_i$; for one-particle states,
  $|\nu_\alpha\rangle = \sum_i U^{*}_{\alpha i}|\nu_i\rangle$.} 
  $\tilde{U}$ (throughout this paper, in our notation the presence of a tilde denotes that a quantity is affected by matter effects). Therefore, after imposing 
the energy conservation condition given by eq.~(\ref{eq:[H,D]}), we get
\be 
\begin{array}{rcl}
H &=&\tilde{U} \,
\text{diag} \left\{\tilde{h}_1,\tilde{h}_2,\tilde{h}_3\right\}
\tilde{U}^\dagger\equiv 
\tilde{U} H_d \tilde{U}^\dagger,
\label{eq:diag}
\\
D_m &=&\tilde{U} \,\text{diag}\left\{d_m^1,d_m^2,d_m^3\right\} \tilde{U}^\dagger\equiv \tilde{U} D_m^d \tilde{U}^\dagger.
\end{array}
\ee
This implies that the average energy is conserved along the whole neutrino propagation (through vacuum and matter). This assumption is
indeed crucial for our analysis. It is expected to be fulfilled in vacuum and in very good approximation in matter. While we assume that 
the quantum decoherence itself does not cause the violation of energy conservation, due to the standard neutrino interaction with matter, for large 
neutrino energies the energy is not exactly conserved in presence of matter due to a small energy transfer to the background fermions. Therefore, in this
case, eq.~(\ref{eq:diag}) does not hold exactly. This issue has been analyzed in detail in \cite{Oliveira:2014jsa,Oliveira:2016asf,Carpio:2017nui}, where it was shown that in a more general
framework in which energy conservation is not assumed, two types of effects in the neutrino oscillation probabilities 
can essentially be distinguished: pure decoherence effects which suppress the oscillating terms, and the so called relaxation effects which affect
non-oscillating terms. 
In~\cite{Morgan:2004vv,Oliveira:2014jsa} it was shown that, for atmospheric neutrino oscillations, the relaxation effects which arise when the energy 
is not conserved are proportional to $\cos^22\theta_{23}$. This suppresses relaxation effects with respect to pure decoherence effects by 
at least two orders of magnitude, since $\cos^2 2\theta_{23}$ is currently constrained by experimental data at the level 
$\cos^2 2\theta_{23} < 0.034$ at 95\% CL~\cite{nufit,Esteban:2016qun}. 
We will thus focus on the analysis of pure decoherence effects in atmospheric neutrino 
oscillations assuming that the neutrino energy is conserved, and therefore eq.~(\ref{eq:diag}) satisfied, along the whole propagation.

From a model-independent point of view, the $d_m^j$ are free parameters that could
a priori depend on the matter effects. However, the most common assumption in the literature is to
assume that the $d_m^j$ are independent of the matter density even in
presence of matter effects.\footnote{This is the case for instance when decoherence is originated by quantum gravity.}. In order to be consistent 
with most previous studies and to compare the bounds obtained in our analysis with the constraints derived in
previous publications, we will also assume that this is the case. Notice that this assumption does not imply that the matter effects are not 
relevant when neutrino propagation is affected by decoherence: it just implies that the $d_m^j$ are assumed to be constant during neutrino propagation 
through the Earth. 

\subsection{Neutrino propagation in uniform matter}
\label{sec:uniform}

Performing the following change of basis 
\be
\tilde{\rho}=\tilde{U}^\dagger\rho\tilde{U},
\ee
eq.~(\ref{eq:1}) can be rewritten as
\begin{align}
\frac{d\tilde{\rho}}{dt} &= -i\left[H_d,\tilde{\rho} \right]-
\sum_m\left[\left\{\tilde{\rho},(D_m^d)^2\right\}-2D_m^d\,\tilde{\rho} \,D_m^{d}\right]\nonumber\\
&-\tilde{U}^\dagger\frac{d\tilde{U}}{dt}\tilde{\rho}-\tilde{\rho}\frac{d\tilde{U}^\dagger}{dt}\tilde{U}.
\label{eq:mass}
\end{align}
If the matter profile is constant along the neutrino path, the system of equations becomes diagonal in $\tilde{\rho}_{ij}$
\be
\frac{d\tilde{\rho}_{ij}}{dt}=-\left[\gamma_{ij}-i\Delta \tilde{h}_{ij}\right]\tilde{\rho}_{ij},
\label{eq:cte}
\ee
where we have defined
\be
\gamma_{ij}\equiv \sum_m\left(d_m^i-d_m^j\right)^2=\gamma_{ji}>0\,;\;\;\;\;\;\;\; \Delta \tilde{h}_{ij}=\tilde{h}_i-\tilde{h}_j.
\label{eq:gamma_dm}
\ee
Therefore, the solution of eq.~(\ref{eq:1}) for constant matter is simply given by
\be
\rho_{\alpha\beta}(t)= \left[\tilde{U}\tilde{\rho}(t)\tilde{U}^\dagger\right]_{\alpha\beta},
\label{sol1}
\ee
with
\be
\tilde{\rho}_{ij}(t)=\tilde{\rho}_{ij}(0)\, e^{-\left[\gamma_{ij}-i\Delta \tilde{h}_{ij}\right]t},
\label{sol2}
\ee
where $\tilde{\rho}_{ij}(0)$ is determined by the initial conditions
of the system. For instance, if the source produces only neutrinos 
of flavor $\alpha$, the initial conditions are given by
\be
\tilde{\rho}_{ij}(0)=\tilde{U}^*_{\alpha i}\tilde{U}_{\alpha j}\, .
\label{init_alpha}
\ee
As a result, the oscillation probabilities in presence of decoherence (for a constant matter profile) read 
\begin{align}
P_{\alpha\beta}& \equiv P(\nu_\alpha \to \nu_\beta) = 
\text{Tr}\left[\hat{\rho}^{(\alpha)}(t)\hat{\rho}^{(\beta)}(0)\right]\nonumber\\
&=
\text{Tr} \left[\hat{\rho}^{(\alpha)}(t)|\nu_\beta\rangle\langle\nu_\beta|\right]=\langle\nu_\beta|\hat{\rho}^{(\alpha)}(t)|\nu_\beta\rangle \nonumber\\
&=\sum_{i,j}\tilde{U}_{\beta i}\tilde{U}^*_{\beta j}\, \tilde{\rho}_{ij}(t)
\nonumber\\
&=\sum_{i,j}\tilde{U}^*_{\alpha i}\tilde{U}_{\beta i}\tilde{U}_{\alpha j}\tilde{U}^*_{\beta j}
e^{-\left[\gamma_{ij}-i\Delta \tilde{h}_{ij}\right]t} \, ,
\end{align}
where $\hat\rho$ denotes the density operator. 
Finally, after some manipulation the above equation can be rewritten in the more familiar form
\begin{align}
P_{\alpha\beta}&=\delta_{\alpha\beta}-2\sum_{i<j}{\rm Re}\left[\tilde{U}_{\alpha i}^*\tilde{U}_{\beta i}
\tilde{U}_{\alpha j}\tilde{U}_{\beta j}^*\right]\left(1-e^{-\gamma_{ij}L}\cos\tilde{\Delta}_{ij}\right)
\nonumber\\
&-2 \sum_{i<j}{\rm Im}\left[\tilde{U}_{\alpha i}^*\tilde{U}_{\beta i}
\tilde{U}_{\alpha j}\tilde{U}_{\beta j}^*\right]e^{-\gamma_{ij}L}\sin\tilde{\Delta}_{ij},
\label{P_alpha_beta}
\end{align}
with
\be
\label{gammaE}
\tilde{\Delta}_{ij}
\equiv\frac{\Delta\tilde{m}^2_{ij}L}{2E},\;\;\;\;\;\;\;\;\;\;
\gamma_{ij}=\gamma_{ji}\equiv\gamma_{ij}^0\left(\frac{E}{\text{GeV}}\right)^n,
\ee 
where $\Delta\tilde{m}^2_{ij} \equiv \tilde{m}^2_i - \tilde{m}^2_j$ are
the effective mass squared differences of neutrinos in matter
and we have used the approximation $L\approx t$, $L$ being 
the distance traveled by the neutrino as it propagates. Note that the power-law dependence
on the neutrino energy given by eq.~(\ref{gammaE}) breaks Lorentz invariance, except for the case with $n=-1$
which gives similar effects to the neutrino decay (see e.g. \cite{GonzalezGarcia:2008ru}).
However, the effect encoded in $\gamma_{ij}$ only suppresses the
oscillatory terms in the oscillation probability while a neutrino decay 
would also affect the non-oscillatory terms. 
Therefore, in the framework considered in this work the total sum of the probabilities adds up to $1$,
while this is not the case for neutrino decay.

From eqs.~(\ref{P_alpha_beta}) and (\ref{gammaE}), one would expect to have a sizable effect in neutrino oscillations for $ \gamma_{ij} L \sim 1$. This condition gives an estimate of the values of $\gamma_{ij}$ for which an effect may be experimentally observable:
\be
\label{gammaL}
\gamma_{ij}^0 \sim 1.7 \cdot 10^{-19} \left(\frac{L}{\text{km}}\right)^{-1}
\left(\frac{E}{\text{GeV}}\right)^{-n} \ \text{GeV}.
\ee
Nevertheless, we would like to remark that fulfilling this condition is not enough to have sensitivity to 
decoherence effects, as we will discuss in the next subsection. 

Even though in our simulations we will numerically compute the exact oscillation probabilities, in order to understand qualitatively the impact of decoherence 
on the oscillation pattern it is useful to derive approximate analytical expressions. In this work, we will be focusing on the study of atmospheric neutrino 
oscillations, for which the oscillation channel $P_{\mu\mu}$ is most relevant. Recently, in~\cite{Denton:2016wmg,Denton:2018hal} approximated but
very accurate analytical expressions for the standard oscillation probabilities in presence of constant matter density were derived. For the 
$\nu_\mu \to \nu_\mu$ oscillation channel including decoherence effects, using the same parametrization as in ref.~\cite{Denton:2018hal}, we find:
\begin{align}
\label{PmumuActe}
P_{\mu\mu}&=1-A_{21} \left[1-e^{-\gamma_{21}L}\cos\tilde{\Delta}_{21}\right]-A_{32} \left[1-e^{-\gamma_{32}L}\cos\tilde{\Delta}_{32}\right]\nonumber\\
&-A_{31} \left[1-e^{-\gamma_{31}L}\cos\tilde{\Delta}_{31}\right],
\end{align}
where
\begin{align}
A_{ij} &\equiv A_{ij}(\theta_{23},\tilde{\theta}_{12},\tilde{\theta}_{13},\delta)\nonumber\\
&= 2|U_{\mu i}(\theta_{23},\tilde{\theta}_{12},\tilde{\theta}_{13},\delta)|^2|U_{\mu j}(\theta_{23},\tilde{\theta}_{12},\tilde{\theta}_{13},\delta)|^2,
\end{align}
and the effective mass splittings and mixing angles in matter can be expressed as~\cite{Denton:2018hal}:
{\small
\begin{align}
\cos2\tilde{\theta}_{13}&=\frac{\cos2\theta_{13}-a/\Delta m^2_{ee}}{\sqrt{(\cos2\theta_{13}-a/\Delta m^2_{ee})^2+\sin^2 2\theta_{13}}},
\nonumber\\
\cos 2\tilde{\theta}_{12}&=\frac{\cos2\theta_{12}-a'/\Delta m^2_{21}}{\sqrt{(\cos2\theta_{12}-a'/\Delta m^2_{21})^2+\sin^2 2\theta_{12}\cos^2(\tilde{\theta}_{13}-\theta_{13})}},
\nonumber\\
\Delta\tilde{m}^2_{21}&=\Delta m^2_{21}\sqrt{(\cos2\theta_{12}-a'/\Delta m^2_{21})^2+\sin^2 2\theta_{12}\cos^2(\tilde{\theta}_{13}-\theta_{13})},
\nonumber\\
\Delta\tilde{m}^2_{31}&=\Delta m^2_{31}+(a-\frac{3}{2}a')+\frac{1}{2}(\Delta \tilde{m}^2_{21}-\Delta m^2_{21}),
\nonumber\\
\Delta\tilde{m}^2_{32}&=\Delta\tilde{m}^2_{31}-\Delta\tilde{m}^2_{21}.
\end{align}}
Here, $a \equiv 2\sqrt{2}G_Fn_e E$, where $G_F$ is the Fermi constant and $n_e$ is the electron number density along the neutrino path,
$\Delta m^2_{ee}\equiv \cos^2\theta_{12}\Delta m^2_{31}+\sin^2\theta_{12}\Delta m^2_{32}$, and 
$a'= a\cos^2\tilde{\theta}_{13}+\Delta m^2_{ee}\sin^2(\tilde{\theta}_{13}-\theta_{13})$. The corresponding 
probability for antineutrinos is obtained simply replacing $a \to -a$
and $\delta \to -\delta$, where $\delta$ denotes the Dirac CP phase.

\subsection{Neutrino propagation in non-uniform matter: adiabatic regime}
\label{sec:adiabatic}

Equation~(\ref{P_alpha_beta}) applies for constant density profiles (which is a very good approximation in the case of long-baseline neutrino oscillation experiments such as T2K or NOvA), but if the matter density is not constant the analysis 
becomes more complicated. Nevertheless, when the adiabaticity condition 
${d \tilde{U}}/{dt}\ll 1$ is fulfilled, as in the solar neutrino case, the solution of the evolution
equations given by eqs.~(\ref{sol1}) and (\ref{sol2}) is still a good
approximation. In such a case, the oscillation probability is given by
\begin{align}
P_{\alpha\beta} &=
\langle\nu_\beta|\hat{\rho}^{(\alpha)}(t)|\nu_\beta\rangle\nonumber\\
&=\sum_{i,j}\tilde{\rho}^{(\alpha)}_{ij}(0)
e^{-\left[\gamma_{ij}-i\Delta \tilde{h}_{ij}\right]t}\langle\nu_\beta|\tilde{\nu}_i^{eff}\rangle\langle\tilde{\nu}_j^{eff}|\nu_\beta\rangle,
\end{align}
where $\nu_i^{eff}$ denotes the effective mass eigenstates
at time $t$. In the case of solar neutrinos, the initial flux of
$\nu_e$ is produced in the solar core and the initial conditions
are given by:
\be
\tilde{\rho}^{(e)}_{ij}(0)=\tilde{U}_{e i}^{0*}\tilde{U}_{e j}^{0},
\ee
where $\tilde{U}^{0}$ denotes the effective mixing matrix at the
production point. On the other hand, since the evolution is adiabatic,
when the neutrinos come out from the Sun we have
$|\tilde{\nu}_i^{eff}\rangle = |\nu_i\rangle$ and thus
\begin{align}
P_{e \beta}& \approx\sum_{i,j} \tilde{U}_{e i}^{0*}U_{\beta i}\tilde{U}^{0}_{e j}U_{\beta j}^*
e^{-\left[\gamma_{ij}-i\Delta \tilde{h}_{ij}\right]t}
\nonumber\\
&= \sum_{i} |\tilde{U}_{e i}^{0}|^2|U_{\beta i}|^2\nonumber
\\
&+2\sum_{i<j}{\rm Re}\left[\tilde{U}_{e i}^{0*}U_{\beta i}
  \tilde{U}_{e j}^{0}U_{\beta j}^*\right]e^{-\gamma_{ij}t}\cos\tilde{\Delta}_{ij}\nonumber\\
&-2 \sum_{i<j}{\rm Im} \left[\tilde{U}_{e i}^{0*}U_{\beta i}
\tilde{U}_{e j}^{0}U_{\beta j}^*\right]e^{-\gamma_{ij}t}\sin\tilde{\Delta}_{ij}.
\end{align}
Finally, for solar neutrinos observed at the Earth we obtain, after averaging over the oscillating phase:
\be
P_{e \beta} \approx \sum_{i} |\tilde{U}_{e i}^{0}|^2|U_{\beta i}|^2,
\ee
which coincides with the standard result. In other words, the
decoherence effects encoded in $\gamma_{ij}$ can not be bounded by
solar neutrino oscillation experiments. This is due to the standard loss of
coherence in the propagation from the Sun to the Earth, which strongly
suppresses the oscillating terms. 
In a similar fashion high-energy astrophysical neutrinos at IceCube are not 
sensitive to decoherence either, since these neutrinos are produced 
in distant astrophysical sources and thus the oscillations will have averaged out by the time they reach the detector.

\subsection{Neutrino propagation in non-uniform matter: layers of constant density}
\label{sec:layers}

In the atmospheric neutrino case, the matter profile cannot be
considered constant since the neutrinos propagate through the Earth
crust, mantle and core, which have different densities. The
adiabaticity condition is not fulfilled either. In this case, eq.~(\ref{eq:mass}) should be solved 
including the non-adiabatic terms, which give non-diagonal
contributions. Even though this can be done numerically, we will show that 
dividing the matter profile into layers of constant density considerably simplifies the
analysis and reduces the computational complexity of the problem. In particular, this is crucial 
in the case of atmospheric neutrino oscillation experiments, for which numerical studies are 
already computationally demanding even in the standard three-family scenario. 
Dividing the matter profile into layers of different constant densities has proved to be a very 
good approximation in the standard three-family scenario and, therefore, we expect the same 
level of accuracy in presence of decoherence. Since the matter is constant in each layer, the
evolution equations can be solved for each layer $M$ as in sec.~\ref{sec:uniform}:
\begin{align}
\rho_{\alpha\beta}^M (t_M) & = \left[ \tilde{U}^{M}\tilde{\rho}^{M}(t_M)(\tilde{U}^M)^\dagger \right]_{\alpha\beta},
\nonumber\\
\tilde{\rho}^{M}_{ij}(t_M)&=\tilde{\rho}^{M}_{ij}(t_{M,0})\, e^{-\left[\gamma_{ij}-i\Delta \tilde{h}^{M}_{ij}\right]\Delta t_M} ,
\label{sol_layer}
\end{align}
where $\Delta t_M \equiv t_M- t_{M,0}$, and $t_{M,0}$ and $t_M$ denote the initial and final times for the propagation along layer $M$, respectively.
Now the problem of computing the probability is just reduced to performing properly the matching among the evolution on the different
layers. Let us first consider the simplest case of two layers $A$ and $B$. The oscillation probability when the neutrino exits the second 
layer (at time $t_B$) is given by
\be
P_{\alpha\beta}= \langle\nu_\beta|\hat{\rho}^{(\alpha)}(t_B)|\nu_\beta\rangle
=\sum_{i,j}\tilde{U}^{B}_{\beta i}\,\tilde{U}^{B*}_{\beta j}\, 
\tilde{\rho}^{B}_{ij}(t_{B,0})\, e^{-\left[\gamma_{ij}-i\Delta \tilde{h}^{B}_{ij}\right]\Delta t_{B} } \, .
\label{Pab_layer1}
\ee
The key point here is that the matching should be done between the solutions of eq.~(\ref{eq:1}) at the frontier between the two
layers and in the flavor basis, as
\be
\rho^{A}_{\alpha\beta}(t_A) = \rho^{B}_{\alpha\beta}(t_{B,0}).
\label{matching}
\ee
After imposing the matching condition, the elements of the density matrix in the second layer at $t_{B,0}$ can be written in the matter basis as:
\begin{align}
\tilde{\rho}^{B}_{ij}(t_{B,0}) &=
\left[(\tilde{U}^{B})^\dagger\tilde{U}^{A}\tilde{\rho}^{A}(t_A)(\tilde{U}^{A})^\dagger \tilde{U}^{B}\right]_{ij}
\nonumber\\
&= \tilde{U}^{B*}_{\delta i}\tilde{U}^{A}_{\delta l}\tilde{\rho}^{A}_{l n}(t_{A,0})
e^{-\left[\gamma_{ln}-i\Delta \tilde{h}^{A}_{ln}\right]\Delta t_A}
\tilde{U}^{A*}_{\gamma n}\tilde{U}^{B}_{\gamma j}
\nonumber\\
&=
\tilde{U}^{B*}_{\delta i}\tilde{U}^{A}_{\delta l}  \tilde{U}^{A*}_{\alpha l} \tilde{U}^{A}_{\alpha n}  
e^{-\left[\gamma_{ln}-i\Delta \tilde{h}^{A}_{ln}\right]\Delta t_A}
\tilde{U}^{A*}_{\gamma n}\tilde{U}^{B}_{\gamma j},
\end{align}
where we have considered that the initial flux is made of
$\nu_\alpha$ as initial condition for the first layer.
After substituting this result into eq.~(\ref{Pab_layer1})
we finally obtain
\begin{align}
P_{\alpha\beta}&= \sum_{\delta,\gamma,i,j,l,n}
\tilde{U}^{B}_{\beta i}\tilde{U}^{B*}_{\delta i}\tilde{U}^{B}_{\gamma j}\tilde{U}^{B*}_{\beta j}
e^{-\left[\gamma_{ij}-i\Delta \tilde{h}^{B}_{ij}\right]\Delta t_B}
 \nonumber\\
 &\times
\tilde{U}^{A}_{\delta l}\tilde{U}^{A*}_{\alpha l}\tilde{U}^{A}_{\alpha n}\tilde{U}^{A*}_{\gamma n}
e^{-\left[\gamma_{ln}-i\Delta \tilde{h}^{A}_{ln}\right]\Delta t_A} \, .
\end{align}
It can be easily checked that, in the limit $\gamma_{ij}\rightarrow 0$, the standard oscillation probability is recovered.
In the three-layer case, following the same procedure we find
\begin{align} 
P_{\alpha\beta} 
& =   
\sum_{\delta,\gamma,\theta,\phi,i,j,l,n,m,k}
\tilde{U}^{C}_{\beta i}\tilde{U}^{C*}_{\delta i}\tilde{U}^{C}_{\gamma j}\tilde{U}^{C*}_{\beta j}
e^{-\left[\gamma_{ij}-i\Delta \tilde{h}^{C}_{ij}\right]\Delta t_C}
\nonumber\\ 
& \times  
\tilde{U}^{B}_{\delta l}\tilde{U}^{B*}_{\theta l}\tilde{U}^{B}_{\phi n}\tilde{U}^{B*}_{\gamma n} 
e^{-\left[\gamma_{ln}-i\Delta\tilde{h}^{B}_{ln}\right]\Delta t_B} 
\nonumber\\ 
& \times 
\tilde{U}^{A}_{\theta m}\tilde{U}^{A*}_{\alpha m}\tilde{U}^{A}_{\alpha k}\tilde{U}^{A*}_{\phi k}
e^{-\left[\gamma_{mk}-i\Delta \tilde{h}^{A}_{mk}\right]\Delta t_A}.  
\label{eq:P-3layers}
\end{align}
The procedure can be easily generalized to an arbitrary number 
of layers. Indeed, under the approximation $L \approx t$, and defining
\be
\tilde{\mathcal{A}}^{M}_{\alpha\beta\gamma\delta} \equiv \sum_{i,j}
\tilde{U}^{M}_{\alpha i}\tilde{U}^{M*}_{\beta i}\tilde{U}^{M}_{\gamma j}\tilde{U}^{M*}_{\delta j}
e^{-\left[\gamma_{ij}-i(\Delta \tilde{m}^{M}_{ij})^2/2E\right] \Delta L_M},
\ee
the probabilities can be written in a more compact way as
for two layers
\be 
P_{\alpha\beta}  = \sum_{\delta,\gamma}
\tilde{\mathcal{A}}^{B}_{\beta\delta\gamma\beta}
\tilde{\mathcal{A}}^{A}_{\delta\alpha\alpha\gamma},
\ee
for three layers
\be
P_{\alpha\beta}  =  \sum_{\delta,\gamma,\theta,\phi}
\tilde{\mathcal{A}}^{C}_{\beta\delta\gamma\beta}
\tilde{\mathcal{A}}^{B}_{\delta\theta\phi\gamma}
\tilde{\mathcal{A}}^{A}_{\theta\alpha\alpha\phi},
\ee
for N layers
\be
P_{\alpha\beta}  = \sum_{\delta,\gamma,\theta,\phi,...,\xi,\omega,\varphi,\rho}
\tilde{\mathcal{A}}^{N}_{\beta\delta\gamma\beta}
\tilde{\mathcal{A}}^{N - 1}_{\delta\theta\phi\gamma} ...\;
\tilde{\mathcal{A}}^{B}_{\xi\varphi\rho\omega}
\tilde{\mathcal{A}}^{A}_{\varphi\alpha\alpha\rho}  \phantom{mmmm}.
\ee

\section{Atmospheric oscillation probabilities with decoherence}
\label{sec:plots-probabilities}

Atmospheric neutrino oscillations take place in a regime where matter effects are significant and can even dominate the oscillations. The relevance 
of matter effect increases with neutrino energy and is very different for neutrinos and antineutrinos, as the sign of the matter potential changes 
between the two cases. Matter effects also depend strongly on the
neutrino mass ordering. 
In order to understand better the numerical results shown in this paper, 
it is useful to derive approximate expressions for the oscillations in the $\nu_\mu\to\nu_\mu$ and $\bar\nu_\mu\to\bar\nu_\mu$ 
channels in the presence of strong matter effects. 

From the results obtained in refs.~\cite{Denton:2016wmg, Denton:2018hal}, 
for neutrino energies $ E \gsim 15$~GeV matter effects drive the effective mixing angles in matter $\tilde\theta_{12}$ and $\tilde\theta_{13}$ to either 0 
or $\pi/2$, depending on the channel (neutrino/antineutrino) and the mass ordering. It is easy to show that, in this regime, the oscillation probabilities 
in eq.~\eqref{PmumuActe} can be approximated as:

\be
P^{\mathrm{NO}}_{\mu\mu} \approx 1 - \frac{1}{2}\sin^22\theta_{23}\left( 1 - e^{-\gamma_{21} L } \cos\tilde \Delta_{21}  \right)
\label{eq:NOnu} 
\ee
for neutrinos, and
\be
P^{\mathrm{NO}}_{\bar\mu\bar\mu} \approx 1 - \frac{1}{2}\sin^22\theta_{23}\left( 1 - e^{-\gamma_{32} L } \cos\tilde \Delta_{32} \right) 
\label{eq:NOnubar}
\ee
for antieneutrinos, assuming a normal ordering (NO). For inverted ordering (IO) we get instead
\be
P^{\mathrm{IO}}_{\mu\mu} \approx 1 -
\frac{1}{2}\sin^22\theta_{23}\left( 1 - e^{-\gamma_{31} L } \cos\tilde
\Delta_{31} \right)
\label{eq:IOnu}
\ee
for neutrinos,
\be
P^{\mathrm{IO}}_{\bar\mu\bar\mu} \approx 1 - \frac{1}{2}\sin^22\theta_{23}\left( 1 - e^{-\gamma_{21} L } \cos\tilde \Delta_{21} \right)
\label{eq:IOnubar}
\ee
for antineutrinos. In the limit $\gamma_{ij} = 0$,
eqs.~\eqref{eq:NOnu}-\eqref{eq:IOnubar} reassemble the standard
neutrino oscillation probabilities derived under the one-dominant
mass-scale approximation~\cite{Pantaleone:1993di}. From
eqs.~\eqref{eq:NOnu}-\eqref{eq:IOnubar} it is easy to see that the
approximated oscillation probabilities for an inverted mass ordering
can be obtained from the corresponding ones for normal mass ordering,
just performing the following transformation:
\begin{align}
\gamma_{21}, \tilde{\Delta}_{21} & \to  \gamma_{31}, \tilde{\Delta}_{31} ,
\label{eq:map1} \\
\gamma_{32}, \tilde{\Delta}_{32} & \to   \gamma_{21}, \tilde{\Delta}_{21} . 
\label{eq:map2}
\end{align}
Moreover, note that since the three decoherence parameters and the three mass splittings are related (see eqs.~(\ref{eq:gamma_dm}) and (\ref{gammaE})), these two transformations 
automatically imply that 
\be
\label{eq:map3}
\gamma_{31},\tilde{\Delta}_{31}  \to  \gamma_{32}, \tilde{\Delta}_{32}. 
\ee

Equations~\eqref{eq:NOnu}-\eqref{eq:IOnubar} illustrate why a proper consideration of the matter effects in the context of three families is of key importance 
in order to correctly interpret the bounds extracted within a simplified two-flavour approximation (as done in e.g. ref.~\cite{Lisi:2000zt}). According to our 
analytical results, which will be confirmed numerically below, the constraints obtained from SK in a two-family approximation cannot be simply applied to $\gamma_{31}$ or $\gamma_{32}$, contrary to the 
naive expectation. In fact, the interpretation of such limits depends strongly on the ordering of neutrino masses and on whether the sensitivity is dominated by the neutrino or antineutrino channels: for neutrinos the decoherence effects at high energies 
would be mainly driven by $\gamma_{21}$ ($\gamma_{31}$) for normal (inverted) ordering. Conversely, in the antineutrino channel
decoherence effects are essentially controlled by $\gamma_{32}$ ($\gamma_{21 }$) for normal (inverted) ordering. Therefore, we conclude that in order to avoid any misinterpretation of the bounds from atmospheric neutrinos, a three-family approach including matter effects should be considered.


\begin{figure*}[h!]
\centering
\includegraphics[width=.95\columnwidth]{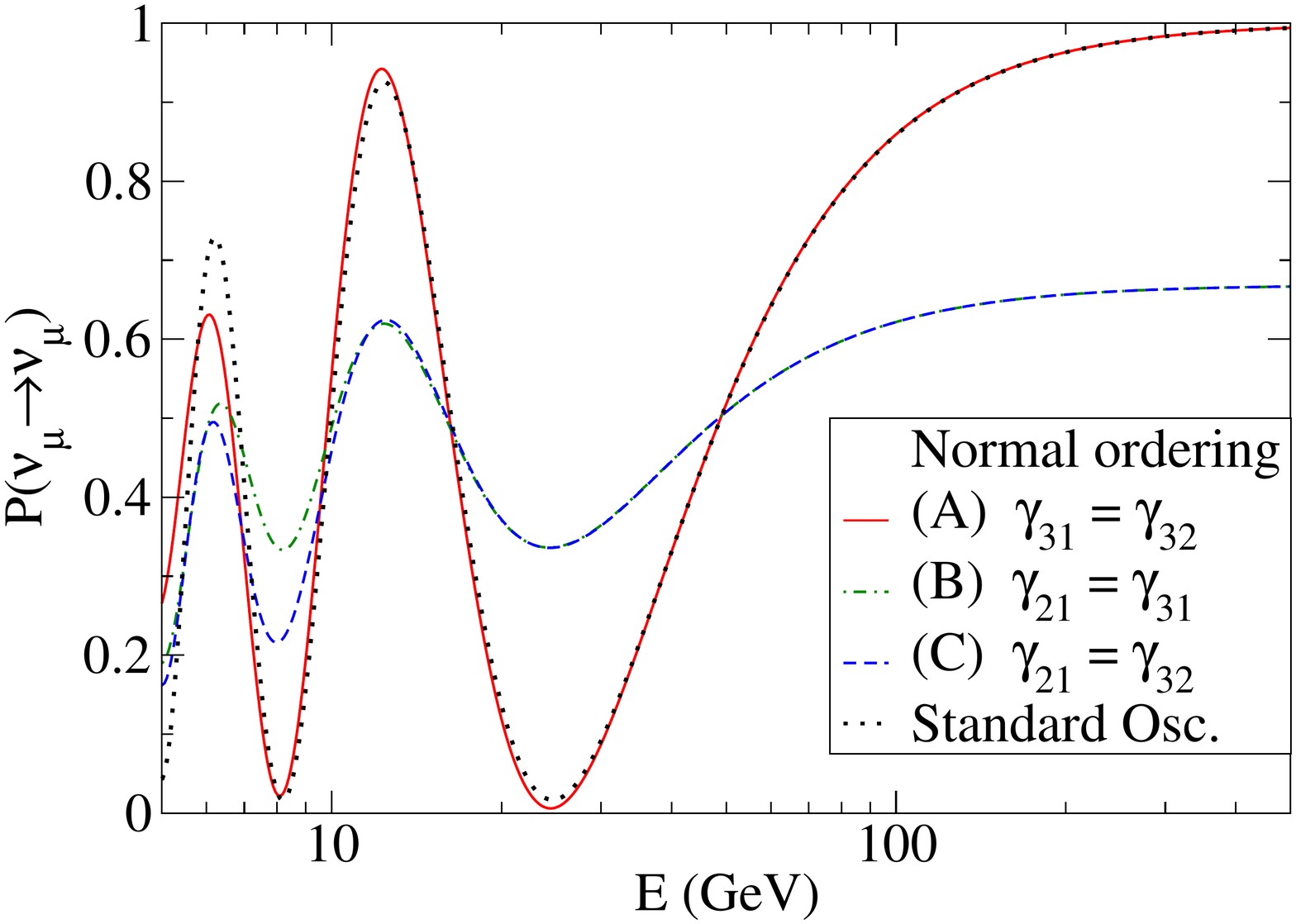}
\includegraphics[width=.95\columnwidth]{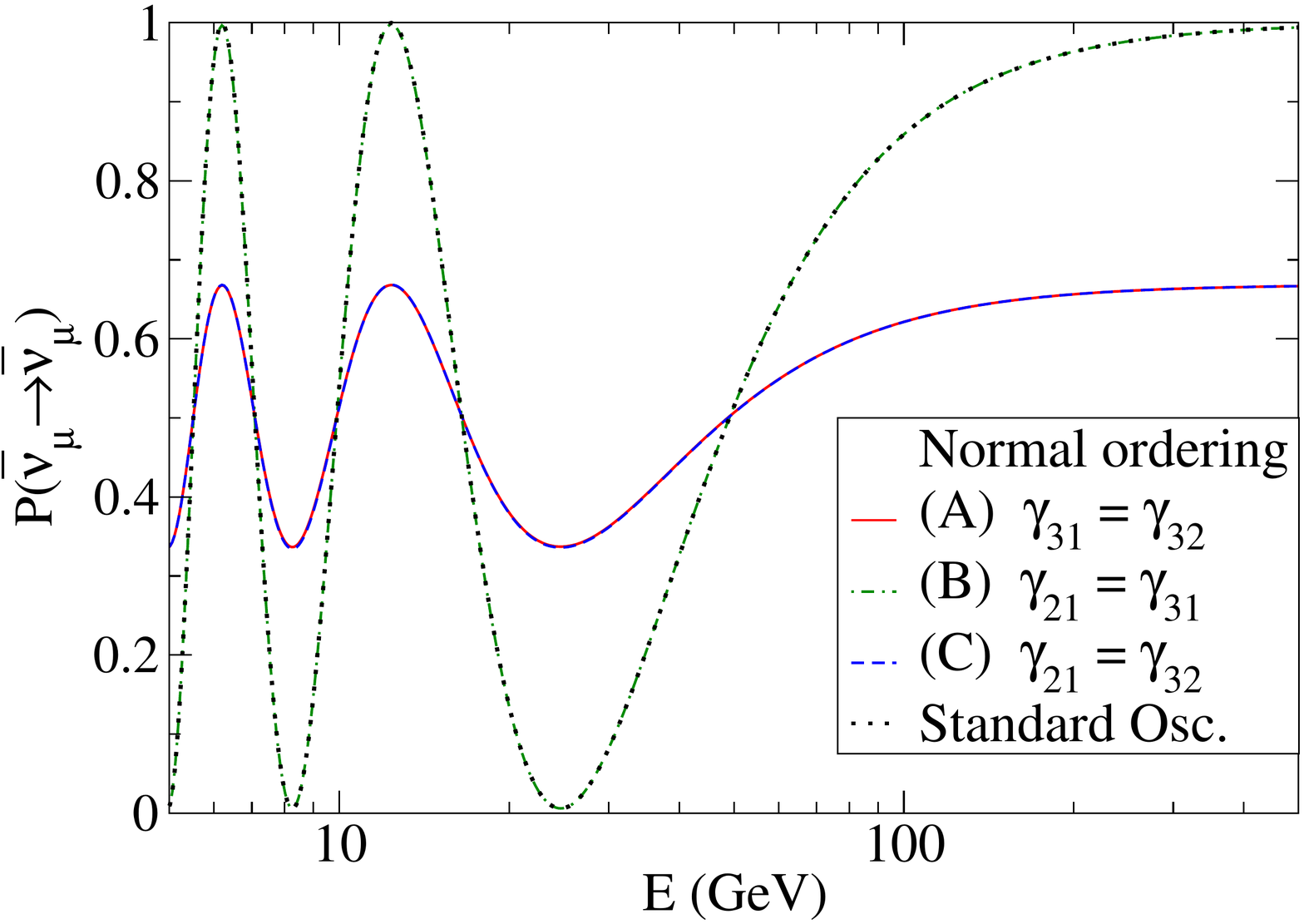}
\includegraphics[width=.95\columnwidth]{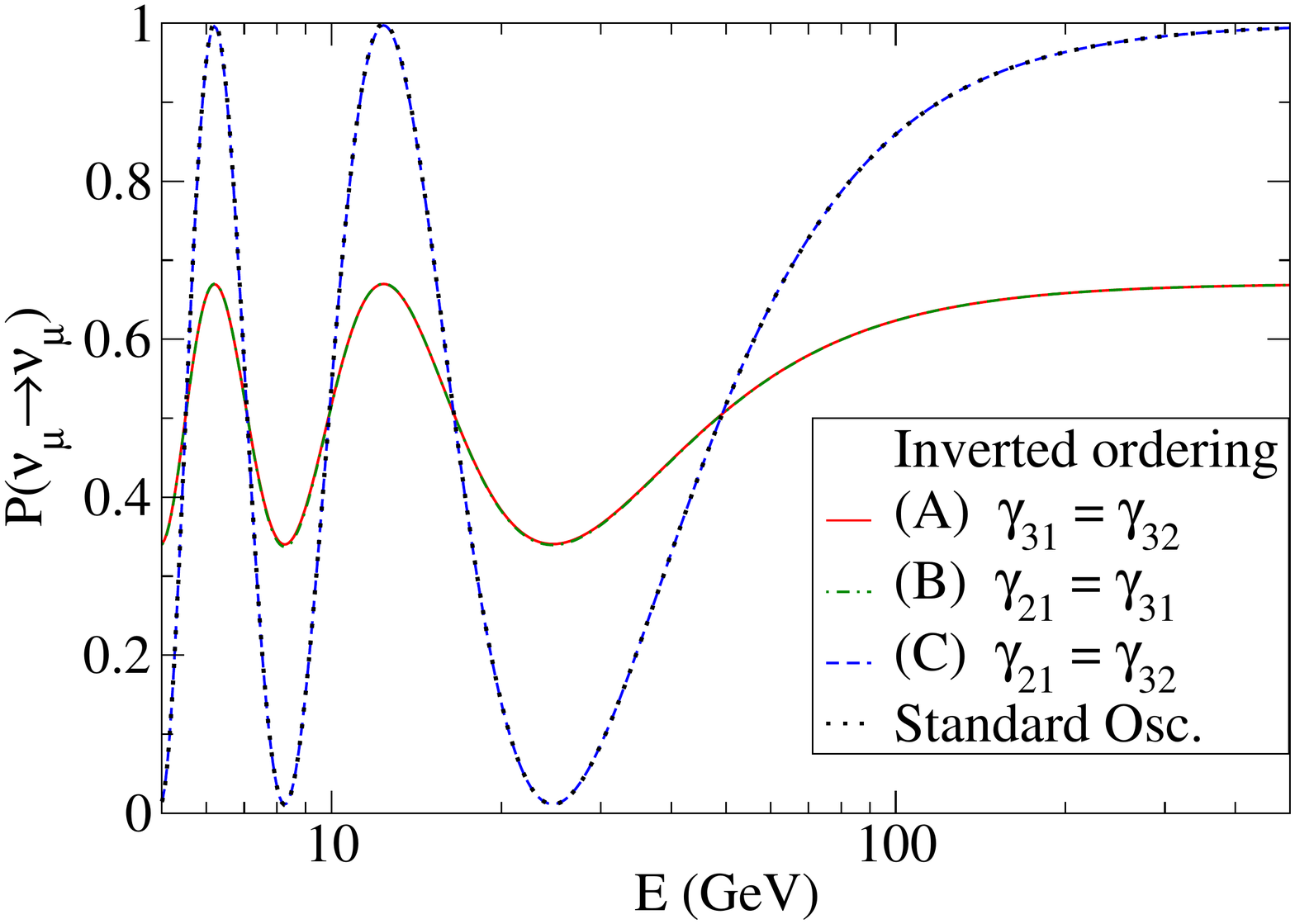}
\includegraphics[width=.95\columnwidth]{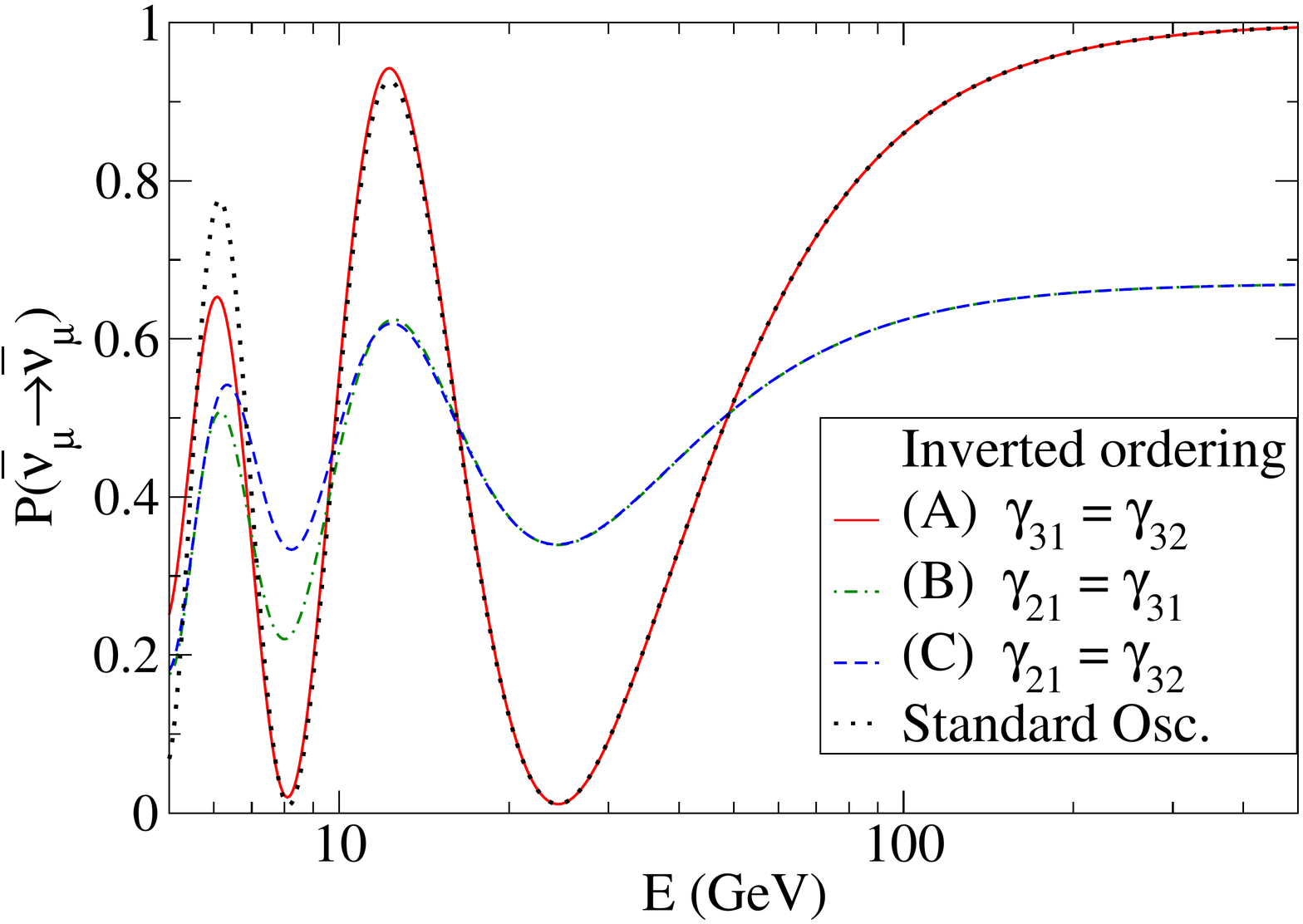}
    \caption{The $\nu_\mu \to \nu_\mu$ (top panels) and $\overline{\nu}_\mu \to \overline{\nu}_\mu$ (bottom panels) oscillation probabilities with ($n=0$) and
without decoherence effects as a function of the neutrino energy. The probabilities have
been computed for normal (left panels) and inverted (right panels) neutrino mass ordering, using a three-layer model for the Earth matter
density profile, and correspond to the case in which the neutrinos cross the center of the Earth. In this figure, in the cases where decoherence effects are included we have set the parameters listed in the legend to the same constant value $\gamma = 2.3 \cdot 10^{-23}$~GeV.
}
\label{fig:prob_mumu}
\end{figure*}

Figure~\ref{fig:prob_mumu} shows the numerically obtained $\nu_\mu \to \nu_\mu$ (top panels) and $\overline{\nu}_\mu \to \overline{\nu}_\mu$ (bottom panels) 
oscillation probabilities for NO (left panels) and IO (right panels), with and without decoherence, as a function of the neutrino energy for a three-layer 
model (details on the accuracy of our three-layer model and the specific parameters used in our simulations can be found in~\ref{app}). For the sake 
of simplicity, in this section we focus on the case $n=0$, where the $\gamma_{ij}$ do not depend on the neutrino energy (the results for different values of 
$n$ show a similar qualitative behavior). The standard oscillation parameters have been fixed to the best-fit values given in~\cite{nufit,Esteban:2016qun}.

Figure~\ref{fig:prob_mumu} clearly shows how the decoherence tends to damp the oscillatory
behavior, in qualitative agreement with eq.~(\ref{PmumuActe}) 
and the corresponding approximate expressions given by 
eqs.~\eqref{eq:NOnu}-\eqref{eq:IOnubar}. 
Note that eq.~\eqref{PmumuActe} has been obtained under several approximations and, 
in particular, considering only one layer with constant matter density. However, we should stress that in our simulations the computation of 
the probability has been done numerically, considering a three-layer matter profile (see~\ref{app} for details).

Since the three $\gamma_{ij}$ are not completely independent from one
another (see eq.~(\ref{eq:gamma_dm})), 
in view of eqs.~\eqref{eq:NOnu}-\eqref{eq:IOnubar} and in order
to simplify the analysis, hereafter we will distinguish three different 
representative cases, where the decoherence effects are dominated by just one parameter:

\begin{enumerate}[label=(\Alph*)]
\item Atmospheric limit: $\gamma_{21}=0$ ($\gamma_{32} = \gamma_{31}$),
\item Solar limit I: $\gamma_{32}=0$ ($\gamma_{21} = \gamma_{31}$),
\item Solar limit II: $\gamma_{31}=0$ ($\gamma_{21} = \gamma_{32}$).
\end{enumerate}

\noindent
In~\ref{sec:d5results}, we will show that the bounds derived in these limits correspond
to the most conservative bounds that can be extracted in the general case. As a reference value for the decoherence parameters, in this
section we have considered $\gamma =   2.3 \cdot 10^{-23}$~GeV, for each of the three limiting cases listed above.

The results in fig.~\ref{fig:prob_mumu} show that, for neutrinos with a NO (top left panel), the impact of decoherence is essentially controlled 
by $\gamma_{21}$, in good agreement with eq.~\eqref{eq:NOnu}: no significant effects are seen in the 
atmospheric limit (A), while a similar impact is obtained in the solar limits I (B) and II (C). Conversely, for IO (top right panel) the effects 
are dominated by $\gamma_{31}$ instead: no effect is observed for the solar limit II (C), while in scenarios (A) and (B) the effect is very similar. 
This can be qualitatively understood from the approximate probability derived in eq.~\eqref{eq:IOnu}, which only depends on the 
decoherence parameter $\gamma_{31}$. On the other hand, in the antineutrino case for NO (bottom left panel) 
no observable decoherence effects take place in case  (B), while cases (A) and (C) show a similar behavior, in agreement with eq.~\eqref{eq:NOnubar}. Conversely, 
for IO (bottom right panel) decoherence effects are essentially controlled by $\gamma_{21}$ as shown in eq.~\eqref{eq:IOnubar}: therefore, no 
significant effects are observed in case (A) while a similar impact is obtained for case (B) and (C). 

\begin{figure*}[ht!]
  \centering
  \includegraphics[width=1.65\columnwidth]{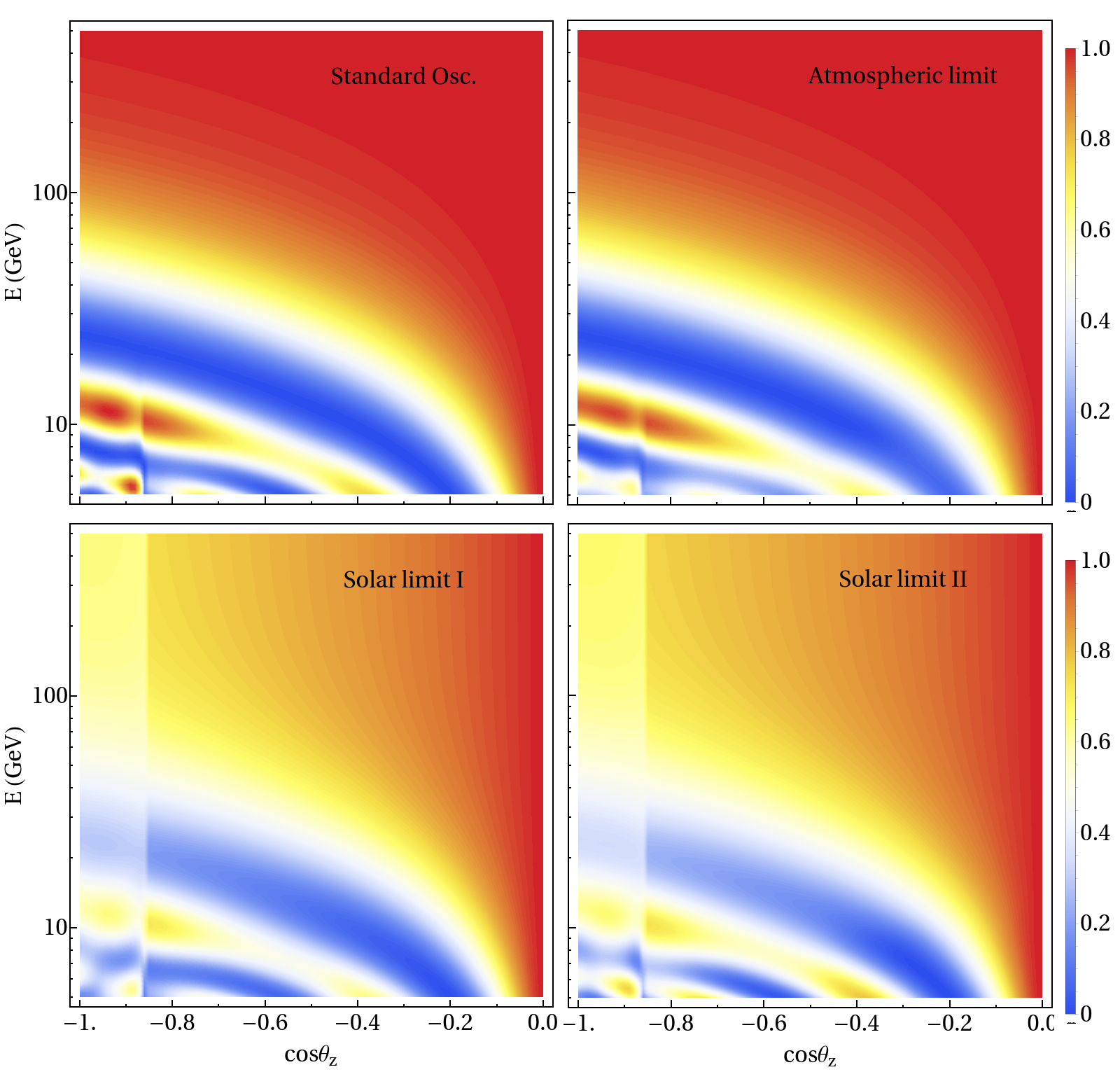}
  \caption{
    Oscillograms for the neutrino oscillation probability $P_{\mu\mu}$, assuming normal mass ordering. 
    The top left panel corresponds to the case of no decoherence
    $\gamma_{ij} = 0$ 
    whereas the rest of the panels correspond to the three limiting cases mentioned in the text:
    (A) 
    $\gamma_{32} = \gamma_{31} $ 
    (top right),  
    (B) 
    $\gamma_{31} = \gamma_{21} $ 
    (bottom left)
    and  (C) 
    $\gamma_{32} = \gamma_{21} $ 
    (bottom right). In all cases, the size of the decoherence parameters that are turned on is set to a constant value, $\gamma =  2.3 \cdot  10^{-23}$~GeV.}
  \label{fig:OscillogramNH}
\end{figure*}

Moreover, it should be pointed out that the transformations listed in Eqs.~\eqref{eq:map1}-\eqref{eq:map3} automatically imply the following equivalence 
for the results obtained in the three limiting cases listed above:
\begin{align}
({\rm A})^{\rm NO} &\longleftrightarrow ({\rm C})^{\rm IO},   \nonumber\\
({\rm B})^{\rm NO} &\longleftrightarrow ({\rm A})^{\rm IO},    \label{eq:map} \\
({\rm C})^{\rm NO} &\longleftrightarrow ({\rm B})^{\rm IO}.   \nonumber
\end{align}
This is also confirmed at the numerical level, as it can be easily seen by
comparing the different lines shown in the left (NO) and right (IO)
panels in fig.~\ref{fig:prob_mumu} for the three limiting cases.

It is also remarkable that, for both normal and inverted mass orderings, even when the standard oscillations turn off (at very high energies), there is 
still a large effect on the probability due to decoherence effects, that could potentially be tested with neutrino telescopes like IceCube. 
In particular, for $E\gsim 200$ GeV one can approximate $\cos\tilde{\Delta}_{ij}\approx 1$, $\forall i,j $. Therefore,
in the standard case (with $\gamma_{ij}=0$) the last three terms in eq.~(\ref{PmumuActe}) approximately vanish, leading to $P_{\mu\mu}\approx1$. However, 
in presence of decoherence those terms will not vanish completely, as $e^{-\gamma_{ij} L}\cos\tilde{\Delta_{ij}} \neq 1$. This
leads to a depletion of $P_{\mu\mu}$, which is no longer equal to 1 in this case. The size of the effect will of course depend on the baseline of 
the experiment. Since at high energies the oscillation probability does no longer depend on the neutrino energy, at oscillation experiments with a
fixed baseline the effect may be hindered by the presence of any systematic errors affecting the normalization of the signal event rates. However, 
at atmospheric experiments this effect can be disentangled from a simple normalization error by comparing the event rates at different nadir angles. 


The dependence of the neutrino probabilities with the zenith
angle $\theta_z$ is illustrated in fig.~\ref{fig:OscillogramNH}, assuming a normal mass ordering and fixing 
the standard oscillation parameters to the best-fit values given in~\cite{nufit,Esteban:2016qun}. The results are shown as a 
neutrino oscillogram (see for instance \cite{Akhmedov:2006hb}), which represents the oscillation probability in the $P_{\mu\mu}$ channel as a function of neutrino energy and zenith angle $\theta_z$ (which can be related to the distance traveled by the neutrino).
Figure~\ref{fig:OscillogramNH} shows the oscillation probability $P_{\mu\mu}$ in the three limiting cases described above,
comparing it to the results in the standard scenario ($\gamma_{ij} = 0$). As expected, the effects depend on the direction of the incoming neutrino and 
they are more relevant in the region $-1\lesssim \cos\theta_z \lesssim -0.4$, this is, for very long 
baselines. This was to be expected, since the decoherence effects are driven by $e^{-\gamma_{ij}L}$. In addition, the dependence of the oscillation 
probability with the zenith angle at very high energies ($ E > 100$~GeV)
is clearly visible in 
the bottom panels of fig.~\ref{fig:OscillogramNH}. 
As we will show in sec.~\ref{sec:results}, 
this will lead to an impressive sensitivity for the IceCube setup.
Finally, note that the results for inverted ordering show similar
features to those in 
fig.~\ref{fig:OscillogramNH}, once the mapping in eq.~(\ref{eq:map}) 
is applied, and are therefore not shown here.

\section{IceCube/DeepCore simulation details and data set }
\label{sec:icecube}

The IceCube neutrino telescope, located at the South Pole, is composed of
5160 DOMs (Digital Optical Module) deployed between 1450~m and 2450~m
below the ice surface along 86 vertical
strings~\cite{Aartsen:2016nxy}. In the inner core of the detector, a
subset of these DOMs were deployed deeper than 1750~m and closer to each
other than in the rest of IceCube. This subset of strings is called
DeepCore. Due to the shorter distance between its DOMs, the neutrino
energy threshold in DeepCore ($\sim 5$~GeV) is lower than in IceCube
($\sim 100$~GeV). This allows DeepCore to observe neutrino events in
the energy region where atmospheric oscillations take place, see
fig.~\ref{fig:prob_mumu}, whereas IceCube only observes high-energy atmospheric neutrino events.

As outlined in sec.~\ref{sec:formalism}, for high energy astrophysical neutrinos
the  effect of non-standard decoherence in the probability would be
completely erased by the time they reach the detector. Therefore, in
this work we will focus on the observation of atmospheric neutrino
events at both IceCube and DeepCore, in the energy range $\sim10$ GeV to
$\sim1$ PeV. In particular, we have used the three-year DeepCore data on
atmospheric neutrinos with energies between $\sim10$ GeV and $\sim1$ TeV, taken between May 2011 and April
2014~\cite{Aartsen:2014yll}, and the one-year IceCube data taken between
2011-2012~\cite{TheIceCube:2016oqi,Jones:2015,Arguelles:2015}, corresponding to neutrinos with energies between 200~GeV and 1~PeV. 

At IceCube and DeepCore, events are divided according to their
topology into ``tracks'' and
``cascades''~\cite{YanezGarza:2014jia}. Tracks are produced by the
Cherenkov radiation of muons propagating in the ice. In atmospheric
neutrino experiments, muons are typically produced by two main
mechanisms: (1) via charged-current (CC) interactions of $\nu_{\mu}$
with nuclei in the detector, and (2) as decay products of mesons
(typically pions and kaons) originated when cosmic rays hit the
atmosphere.
Conversely, cascades are created in CC interactions of
$\nu_{e}$ or $\nu_\tau$\footnote{Technically, a CC $\nu_{\tau}$ event
  could be distinguished from a $\nu_e$ CC event, e.g., by the
  observation of two separates cascades connected by a track from the
  $\tau$ propagation~\cite{Aartsen:2015dlt}. However, for atmospheric
  neutrino energies the distance between the cascades cannot be
  resolved by the DOMs at IceCube/DeepCore, leaving in the detector a
  signal similar to a single cascade. }: in this case, the
rapid energy loss of electrons as they move through the ice is the
origin of an electromagnetic shower. At IceCube/DeepCore, cascades are
also observed as the product of hadronic showers generated in neutral-current 
(NC) interactions for neutrinos of all flavors. 
Our analysis considers only track-like events observed at both IceCube and DeepCore although, as we will see, some small contamination from cascade events can be expected (especially at low energies).

\subsection{IceCube simulation details}

For IceCube, the observed event rates are provided in a grid of $10 \times
21$ bins~\cite{Jones:2015}, using 10 bins for the reconstructed energy
(logarithmically spaced, ranging from 400 GeV to 20 TeV), and 21 bins
for the reconstructed neutrino direction (linearly spaced, between
$\cos\theta_z^{rec} = -1.02$ and $\cos\theta_{z}^{rec} = 0.24$). 
The muon energy is
reconstructed with an energy resolution
$\sigma_{\text{log}_{10}(E_{\mu}/\text{GeV})}\sim
0.5$~\cite{TheIceCube:2016oqi}, while the zenith angle resolution 
is in the range $\sigma_{\cos\theta_{z}}\in [0.005,0.015]$,
depending on the scattering muon angle.

\begin{figure*}[h!]
  \centering
  \includegraphics[width=1.85\columnwidth]{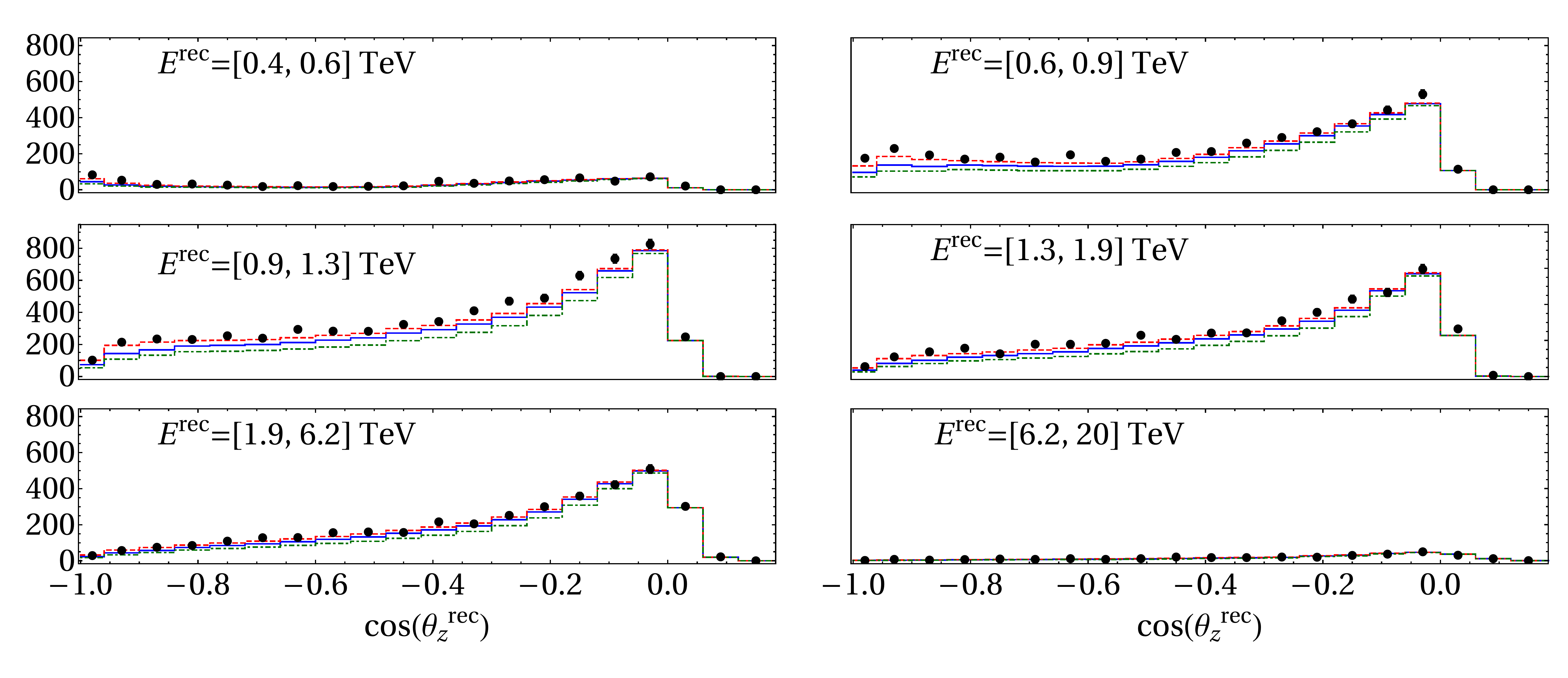}
  \caption{
    Event distributions obtained for IceCube in our numerical simulations as a function of the reconstructed value of 
    the cosine of the zenith angle, for neutrinos in different reconstructed energy ranges. The lines have been obtained assuming a normal mass ordering, for the following values for the decoherence parameters: 
    $\gamma_{21} = \gamma_{31} = 2.3 \cdot 10^{-23}$ GeV (solid blue line), $\gamma_{21} = \gamma_{31} = 10^{-22}$ GeV (dot-dashed green line) and 
    without decoherence (dashed red line). The observed data points~\cite{Jones:2015} are represented by the black dots, and the error bars  indicate the statistical uncertainties for each bin.  
  }
  \label{fig:event-distributionsIC}
\end{figure*}

The number of events in each bin is computed as:
\begin{align}
N_{i} (E^{rec}, \theta_z^{rec})=  & \sum_{\pm}  \int dE \; d\cos\theta_z  \,
\phi^\text{atm}_{\mu,\pm}(E,\theta_z )  P_{\mu\mu}^\pm(E,\theta_z )\nonumber      \label{eq:events-IC} \\
& A^\text{eff}_{i, \pm, \mu}(E,\theta_z ,E^{rec},\theta_z^{rec}) e^{-X(\theta_z)\sigma^{\pm}(E)},
\end{align}
where $E,\theta_z$ denote the true values of energy and zenith angle, 
while $E^{rec},\theta_z^{rec}$ refer to their reconstructed quantities. 
Here, $\phi^\text{atm}_{\mu,\pm}$ is the atmospheric flux for muon
neutrinos~(+) and anti-neutrinos~(-), $P_{\mu\mu}^\pm(E,\theta_z)$
is the neutrino/antineutrino oscillation probability given by
eq.~(\ref{eq:P-3layers}), and $A^\text{eff}_{i,\pm,\mu}(E,\theta_z)$
is the effective area encoding the detector response in neutrino
energy and direction (which relates true and reconstructed
variables), the interaction cross section and a normalization
constant, and has been integrated over the whole data taking period. In our IceCube
simulations, we have used the unpropagated atmospheric flux
(HondaGaisser) provided by the
collaboration~\cite{TheIceCube:2016oqi,IceCubeweb}, and for the
effective area we have used the nominal detector response 
from refs.~\cite{TheIceCube:2016oqi,IceCubeweb}. On the other hand,
the exponential factor takes into account the absorption of the
neutrino flux by the Earth, which increases with the neutrino
energy.  Here, $X(\theta_z)$ is the column density along the
neutrino path and $\sigma^{\pm}(E)$ is the total inclusive cross
section for $\nu_{\mu}$ or $\bar\nu_\mu$. Note that in
eq.~\eqref{eq:events-IC} no contamination from cascade events is
considered since the mis-identification rate is expected to be
negligible at these energies~\cite{Weaver:2015}. Similarly, the
number of atmospheric muons that pass the selection cuts can also be
neglected, given the extremely good angular resolution at these
energies~\cite{TheIceCube:2016oqi}.


Figure~\ref{fig:event-distributionsIC} shows the expected number of events
for IceCube from our numerical simulations including decoherence, for $\gamma_{21} = \gamma_{31} = 2.3 \cdot 10^{-23}$ GeV
 (solid blue lines) and $\gamma_{21} = \gamma_{31} = 10^{-22}$ GeV
 (dot-dashed green lines), as a function of $\cos \theta_z^{rec}$, for events in
 different reconstructed energy ranges. For simplicity, we have considered the $n=0$ case (that is, $\gamma_{ij}$ independent of the neutrino energy). 
The expected result without decoherence is also shown for comparison (dashed red lines), while the observed data~\cite{Jones:2015} are shown by the black dots.

For the analysis of the IceCube data we have performed a Poissonian
log-likelihood analysis doing a simultaneous fit on the following
parameters: $\Delta m^2_{32},\, \theta_{23}$ and $\gamma_{ij}$. The rest of the oscillation parameters have been
kept fixed to their current best-fit values from ref.~\cite{nufit, Esteban:2016qun}. The most relevant systematic errors used in
the fit are summarized in tab.~\ref{tab:sys-IC}, and have been taken
from refs.~\cite{TheIceCube:2016oqi,Arguelles:2015,IceCubeweb}. For each systematic
uncertainty a pull term is added to the $\chi^2$ following the values
listed in the table, except in the cases indicated as ``Free'' (when
the corresponding nuisance parameter is allowed to float freely in the
fit).
 
\begin{table}
\begin{center}
\caption{The most relevant systematic errors used in our analysis of
  IceCube data, taken from
  refs.~\cite{TheIceCube:2016oqi,Arguelles:2015,IceCubeweb}.\label{tab:sys-IC} }
\begin{tabular}{l c }
\hline\noalign{\smallskip}
Source of uncertainty & Value \\
\noalign{\smallskip}\hline\noalign{\smallskip}
Flux - normalization & Free \\
Flux - $\pi/K$ ratio & 10\% \\
Flux - energy dependence as $(E/E_0)^\eta $  & $\Delta \eta = 0.05$ \\
Flux - $\bar{\nu}/\nu$ & 2.5\%\\
DOM efficiency & 5\% \\
Photon scattering & 10\% \\
Photon absorption & 10\% \\
\noalign{\smallskip}\hline
\end{tabular}
\end{center}
\end{table}  

\subsection{DeepCore simulation details}
  
In the case of DeepCore, the observed event rates~\cite{Aartsen:2014yll} 
are provided in a grid of $8\times 8$ bins, using 8 bins for the
reconstructed neutrino energy and 8 bins for the reconstructed
neutrino direction. The energy resolution
  $\sigma_{E/\text{GeV}}$ is in the range of 30\%-20\% while the zenith
  angle resolution improves with the energy, from
  $\sigma_{\theta_{z}} = 12^{\circ}$ at $E_{\nu} = 10$~GeV to $\sigma_{\theta_{z}} = 5^{\circ}$ at $E_{\nu} =
  40$~GeV~\cite{Aartsen:2014yll}. 
  
\begin{figure*}[ht!]
\centering
  \includegraphics[width=1.78\columnwidth]{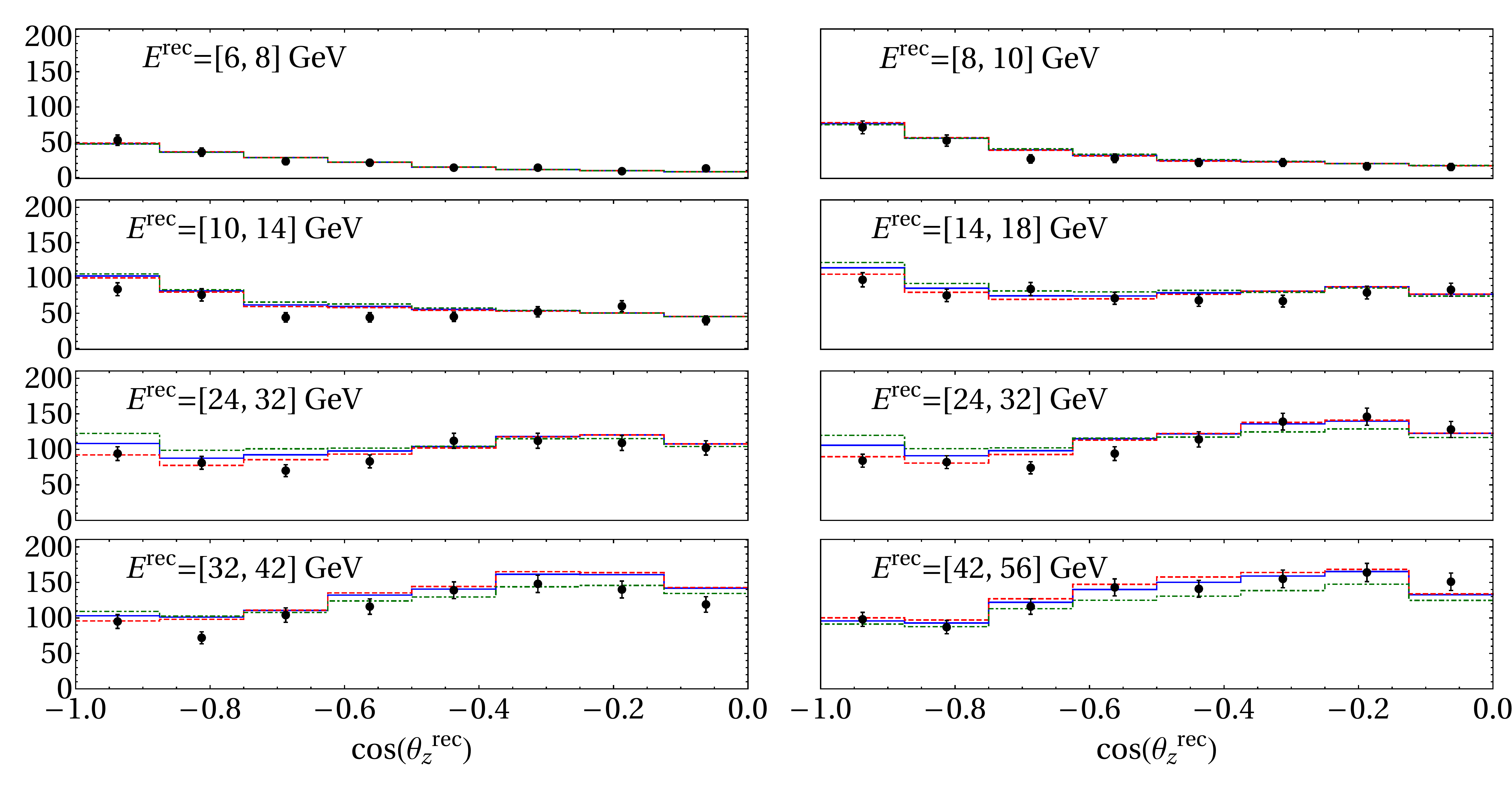}
  \caption{
Event distributions obtained for DeepCore in our numerical simulations as a function of the reconstructed values of the cosine of the 
zenith angle, for neutrinos in different reconstructed energy ranges. The lines have been obtained assuming a normal mass ordering, for the following values for the decoherence parameters: 
$\gamma_{21} = \gamma_{31} = 2.3 \cdot 10^{-23}$ GeV (solid blue line), $\gamma_{21} = \gamma_{31} = 10^{-22}$ GeV (dot-dashed green line) and 
without decoherence (dashed red line). The observed data points~\cite{Aartsen:2014yll} are represented by the black dots, and the error bars indicate the statistical uncertainties for each bin.    
}
  \label{fig:event-distributionsDC}
\end{figure*}
  
In each bin, the number of events is computed as
\begin{align}
  N_{i} (E^{rec},\theta_z^{rec}) &=  
   \sum_{\pm,\alpha,\beta} \int dE\, d\cos\theta_{z} \, 
   \phi^\text{atm}_{\alpha,\pm}(E,\theta_{z}) \, P_{\alpha\beta}^\pm(E,\theta_{z})\nonumber\\
   & A^\text{eff}_{i, \pm, \beta}(E,\theta_{z}, E^{rec},\theta_z^{rec})  \nonumber \\ 
   &+N_{i,\mu} (E^{rec}, \theta_z^{rec}).
  \label{eq:events-DC}
\end{align}
Unlike for IceCube, at DeepCore muon tracks can be produced from
$\nu_\mu \to \nu_\mu$ and $\nu_e \to \nu_\mu$ events\footnote{The flux
  from $\nu_\tau$ can be considered negligible at these
  energies.}. Moreover, the track-like event distributions at DeepCore
will also receive partial contributions from cascades which are
mis-identified as tracks: hence the sum over $\beta = e, \mu,\tau$ in
eq.~\eqref{eq:events-DC}. Therefore, here
$\phi^\text{atm}_{\alpha,\pm}$ stands for the atmospheric flux for
neutrinos/antineutrinos of flavor $\alpha$ (where we have used the
fluxes from ref.~\cite{Honda:2015fha}), and $ P_{\alpha\beta}^{\pm}$
refers to the neutrino/antineutrino oscillation probability in the
channel $\nu_\alpha \to \nu_\beta$ for neutrinos (+) 
(or $ \bar\nu_\alpha \to\bar\nu_\beta$, for antineutrinos (-)). The rejection efficiencies for the
contamination are included in the detector response function
$A^\text{eff}_{i, \pm, \beta}$, which now depends on the flavor
$\beta$ of the interacting neutrino. Finally, an estimate of the
atmospheric muons that overcome the selection criteria (taken from
refs.~\cite{Aartsen:2014yll,IceCubeweb}) is also added for each bin in
reconstructed variables, $N_{i,\mu}$.

Figure~\ref{fig:event-distributionsDC} shows the expected number of events
for DeepCore obtained from our numerical simulations including decoherence, for $\gamma_{21} = \gamma_{31} = 2.3 \cdot 10^{-23}$ GeV (solid blue lines) and $\gamma_{21} = \gamma_{31} = 10^{-22}$ GeV
 (dot-dashed green lines), as a function of $\cos \theta_z^{rec}$, for events in
 different reconstructed energy ranges. For simplicity, we have considered the $n=0$
 case (that is, $\gamma_{ij}$ independent of the neutrino energy). 
The expected result without decoherence is also shown for comparison (dashed red lines), 
while the observed data~\cite{Aartsen:2014yll} are shown by the black dots. 

In this work a Gaussian maximum likelihood is used to analyze the
DeepCore data, performing a simultaneous fit on the following
parameters: $\Delta m^2_{32},\, \theta_{23}$ and
  $\gamma_{ij}$. The rest of the oscillation parameters have been
kept fixed to their current best-fit values from
refs.~\cite{nufit, Esteban:2016qun}. The systematics used in the fit are those
associated with the flux, the detector response and the atmospheric
muons given in ref.~\cite{Aartsen:2014yll} and are summarized in
tab.~\ref{tab:sys-DC}. For each systematic uncertainty a pull term is
added to the $\chi^2$ following the values listed in the table, except
in the cases indicated as ``Free'' (when the corresponding nuisance
parameter is allowed to float freely in the fit). We have checked that
our analysis reproduces the confidence regions in the $\Delta m_{32}^2 -
\theta_{23}$ plane obtained by the DeepCore collaboration in
ref.~\cite{Aartsen:2014yll} to a very good level of accuracy.

\begin{table}
\begin{center}
\caption{Systematic errors used in our analysis of DeepCore data,
  taken from
  refs.~\cite{Aartsen:2014yll,YanezGarza:2014jia}. \label{tab:sys-DC} }
\begin{tabular}{l c }
\hline\noalign{\smallskip}
Source of uncertainty & Value \\
\noalign{\smallskip}\hline\noalign{\smallskip}
Flux - normalization & Free \\
Flux - energy dependence as $(E/E_0)^\eta $  & $\Delta \eta = 0.05$ \\
Flux - $(\nu_{e}+\bar{\nu}_e)/(\nu_{\mu}+\bar{\nu}_{\mu})$ ratio &20\%\\
Background - normalization & Free\\
DOM efficiency & 10\% \\
Optical properties of the ice & 1\%\\
\noalign{\smallskip}\hline
\end{tabular}
\end{center}
\end{table}  
  
  Finally, it should be noted that our fit does not include the latest 
  atmospheric data recently published by the DeepCore collaboration~\cite{Aartsen:2017nmd}. The new
  analysis uses a different data set (from April 2012 to May 2015) and
  a new implementation of systematic errors, which lead to smaller confidence regions in the 
  $\Delta m^2_{32} - \theta_{23}$ plane. However, the detector response parameters and
  systematic errors used in the latest release have not been published yet. 
  In view of the better results obtained for the standard three-family oscillation scenario, a similar improvement is to be expected 
  if the analysis performed in this work were to be repeated using the latest DeepCore data.

\section{Results}
\label{sec:results}

Following the procedure described in sec.~\ref{sec:icecube} we have obtained the
$\chi^2$ for every point in the parameter space. 
Marginalizing over the relevant mixing and mass parameters, 
namely, $\Delta m^2_{32}$ and $\theta_{23}$, the sensitivity of
the data to $\gamma_{ij}$ parameters is determined by evaluating 
the $\sqrt{\Delta\chi^2}$, 
with $\Delta\chi^2 \equiv \chi^2 - \chi^2_\text{min}$, 
where $\chi^2_{\text{min}}$ is the value at 
the global minimum. 

In this section we will only show the results obtained for NO, since we have
checked that extremely similar results are obtained for IO after applying 
the mapping given in eq.~(\ref{eq:map}). Nevertheless, in sec.~\ref{sec:conclusions} we will also provide 
the 95\% confidence level (CL) bounds obtained in our numerical analysis for the IO case. The bounds obtained are in very good 
agreement with the mapping given in eq.~(\ref{eq:map}).

\begin{figure*}[ht!]
\centering
   \includegraphics[scale = 0.3]{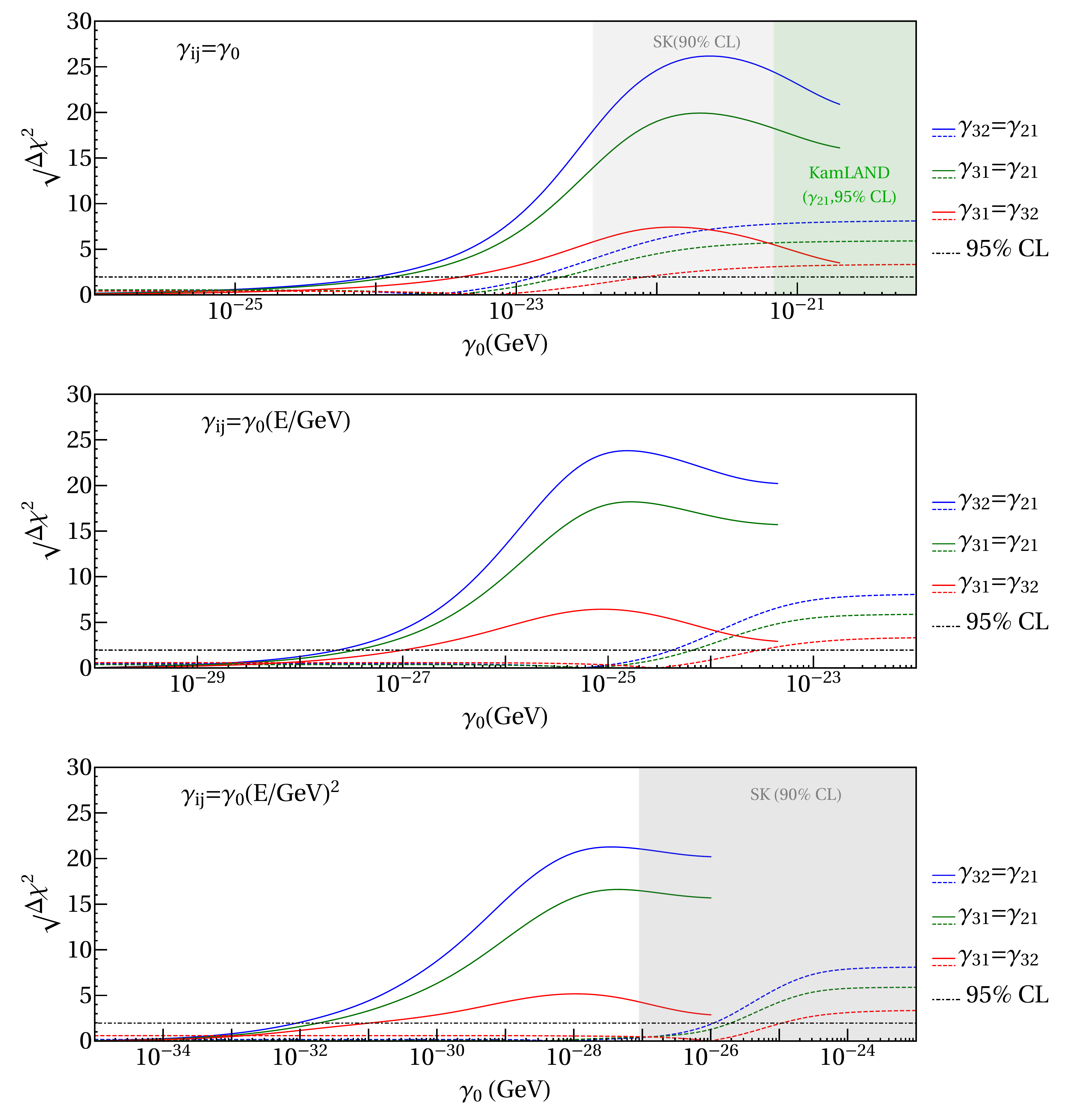}
  \caption{
Values of the $\sqrt{\Delta \chi^2}$ as a function of the decoherence parameter
for the Atmospheric limit (red), Solar limit
 I (green) and Solar limit II (blue) defined in sec.~\ref{sec:plots-probabilities}. 
The results obtained from our analysis of IceCube (DeepCore) data are denoted by the solid (dashed) lines. The three panels have been obtained for NO, assuming a different dependence on the neutrino energy: $n=0$ (top
 panel), $n=1$ (middle panel) and $n=2$ (bottom panel). 
 The shaded regions are disfavored by previous analysis of 
SK~\cite{Lisi:2000zt} and KamLAND~\cite{Gomes:2016ixi} data, see text for details. The horizontal black line indicates the value 
of the $\sqrt{\Delta \chi^2}$ corresponding to 95\% CL for 1 degree of freedom.
}
\label{fig:chi}
\end{figure*}

Figure~\ref{fig:chi} shows the obtained $\sqrt{\Delta \chi^2}$ 
as a function of $\gamma_0$ for the three
limiting cases defined in sec.~\ref{sec:plots-probabilities}: 
(A) atmospheric limit, $\gamma_0 = \gamma_{32}^0=\gamma_{31}^0$ (red curve); 
(B) solar limit I, $\gamma_0 = \gamma_{21}^0=\gamma_{31}^0$  (green curve); and (C) solar limit II, $\gamma_0 = \gamma_{21}^0=\gamma_{32}^0$ (blue curve). In all cases, the solid (dashed) lines
correspond to the results obtained from our analysis of the IceCube (DeepCore) data, and each panel shows the results obtained assuming a different energy 
dependence for the decoherence parameters, see eq.~\eqref{gammaE}:
$n=0$ (top panel), $n=1$ (middle panel) and $n=2$ (bottom panel). 
The shaded regions are disfavored by previous analysis of 
SK~\cite{Lisi:2000zt} (90\% CL) and KamLAND~\cite{Gomes:2016ixi} data (95\% CL).
As explained in sec.~\ref{sec:plots-probabilities}, the KamLAND constraints 
derived in~\cite{Gomes:2016ixi} apply to $\gamma_{12}^0$ (solar limits) while it is not clear to which 
$\gamma_{ij}$ the bounds from SK obtained in~\cite{Lisi:2000zt} would apply, since this depends on the true neutrino mass ordering (which is yet unknown). 

Note that the size of the atmosphere has been neglected in our 
calculations (see~\ref{app} for details). This is a good approximation for small values of the decoherence parameters, but it starts to fail 
if the decoherence effects are large enough to affect neutrinos with $\cos\theta_z > 0$. Therefore, in the case of IceCube we have shown our 
results only in the region where this approximation holds. In the case of DeepCore, due to the smaller energies considered, our approximation has 
no impact on the results even for large values of the decoherence parameters. Therefore, the approximation has only an impact on the IceCube 
results in a region of the decoherence parameter space which is already ruled out either by DeepCore or other experiments.

Figure~\ref{fig:chi} shows that for both DeepCore and IceCube the best
sensitivity is achieved for the solar limits (B) and (C) while the weakest limit 
is obtained in the atmospheric limit (A). In particular, the strongest
bound is obtained for (C). 
This is in agreement
with the behaviour of the oscillation probability in presence of strong matter effects, discussed in
sec.~\ref{sec:plots-probabilities}. 
On one hand, as shown in sec.~\ref{sec:plots-probabilities}, for NO the
decoherence effects are mainly driven by
$\gamma_{21}$ in the neutrino channel and $\gamma_{32}$ in the antineutrino channel. On the other hand, the number of antineutrino events is going to be suppressed with respect to the neutrino case, due
to the smaller cross section and flux. 
Hence, the best sensitivity is expected for case (C), where $\gamma_0 =\gamma_{21}^0=\gamma_{32}^0$, since both 
neutrinos and antineutrinos are sensitive to decoherence effects. Conversely, in case (B), where $\gamma_0 = \gamma_{21}^0=\gamma_{31}^0$, only neutrinos are sensitive to decoherence effects, and therefore some sensitivity is lost with respect to the results for case (C). Finally, in case (A), with $\gamma_0 = \gamma_{32}^0=\gamma_{31}^0$, the bounds come mainly from the impact of decoherence on the antineutrino event rates and, since these are much smaller than in the neutrino case, the obtained bounds are much weaker when compared to the results obtained in case (B).

Figure~\ref{fig:chi} shows a flat asymptotic feature of the DeepCore's $\sqrt{\Delta\chi^2}$
for large values of $\gamma_0$, where the sensitivity becomes independent of $\gamma_0$.
Conversely, for IceCube there is a decrease in sensitivity for values of $\gamma$ above a certain range: for example, for $n=0$ the best sensitivity is achieved 
for $\gamma_0 \sim \mathcal{O}(10^{-22})$~GeV while it decreases for higher values. This behaviour can be understood as follows. 
For the neutrino energies observed at IceCube (above 100~GeV) the oscillation phases do not develop and the probabilities 
do not depend on the energy ($\cos\tilde{\Delta}_{ij}\approx 1$ in eq.~(\ref{PmumuActe})). Therefore, at IceCube the sensitivity to the decoherence effects comes from the observation of a non-standard behaviour of 
the number of events with the zenith angle alone. Naively, eq.~(\ref{gammaL}) gives the values of $L$ and $\gamma$ that yield a large effect. Considering
$n=0$, for example, where there is a one-to-one relation between the two, we get that for $\gamma_0 \sim 10^{-22}$~GeV the 
effect will be maximal for distances of the order $L\sim \mathcal{O}(10^{3})$~km. This is the typical distance traveled by atmospheric neutrinos 
crossing the Earth and therefore the sensitivity of IceCube is maximized in this range. Conversely, for larger
(smaller) values of $\gamma_{0}$, only neutrinos coming from the most horizontal (vertical)
directions are affected, leading to a reduced impact on the $\chi^2$. 

\begin{figure}[t!]
  \centering
 \includegraphics[width=0.98\columnwidth]{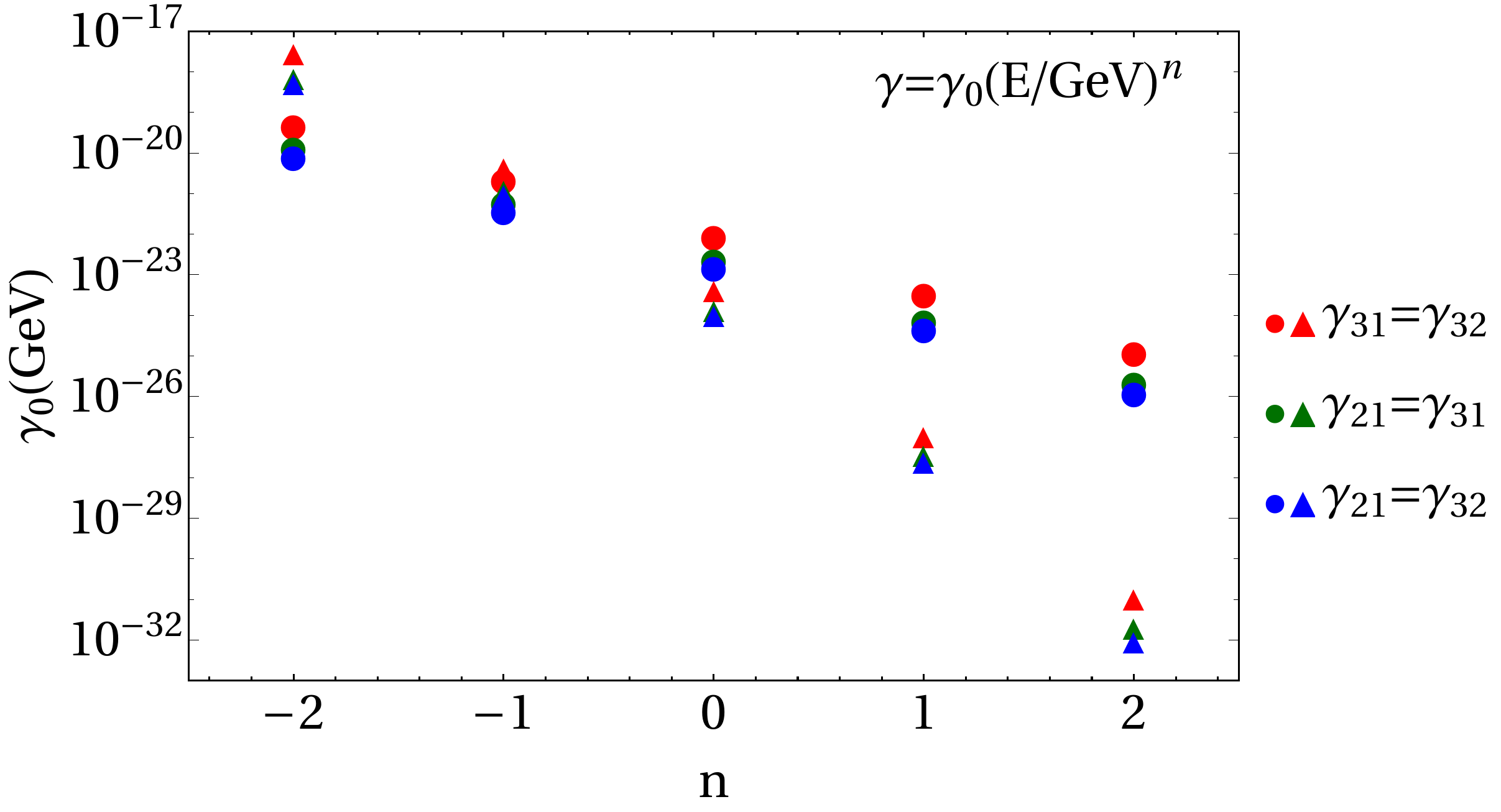}
  \caption{
95\% CL bounds on the decoherence parameters $\gamma_0$, for NO, as a (discrete)
 function of the power-law index $n$ for the Atmospheric limit
 (red), 
Solar limit I (green) and Solar limit II (blue). The solid circles (triangles)
correspond to the DeepCore (IceCube) analysis.}
\label{fig:Bounds-vs-n}
\end{figure}

From the comparison between the different panels in fig.~\ref{fig:chi} we can see that the limits change considerably with the value of $n$, which parametrizes the energy dependence of the decoherence parameters (see eq.~\eqref{gammaE}). In particular, we observe in fig.~\ref{fig:chi} that the sensitivity improves with $n$ and that, as it is increased, the results for IceCube improve much faster (compared to DeepCore) due to the much higher neutrino energies considered in this case. The behaviour of the sensitivities with the value of $n$ is better appreciated in fig.~\ref{fig:Bounds-vs-n}, where we show the bounds obtained at 95\% CL (for 1 degree of freedom) as a
(discrete) function of the power-law index $n$, for $n = -2, -1, 0, 1$ and 2. 
The DeepCore bounds are represented by solid circles while the IceCube constraints are given by the solid triangles. 
The results seem to follow the linear relation
\be
\label{eq:gammavsn}
\ln(\gamma_0 / \text{GeV})={\rm constant} - n \ln (E_0/\text{GeV}),
\ee
where $E_0 \simeq$ 2.5 TeV (30 GeV) for IceCube (DeepCore).
This can be understood as follows. 
Decoherence effects enter the oscillation probabilities only through the factor 
$\gamma L = \gamma_0 (E/\text{GeV})^n L$, for any value of $n$.
Naively, we expect that the sensitivity limit is obtained for $\gamma L \sim \mathcal{O}(1)$ (although 
the precise value will eventually depend on the neutrino mass ordering, on the particular 
$\gamma_{ij}$ which drives the sensitivity, and on the data set considered). 
Taking the logarithm of $\gamma_0 (E/\text{GeV})^n L$ = constant, 
we reproduce eq.~(\ref{eq:gammavsn}). At first approximation, 
the value of $E_0$ in eq.~\eqref{eq:gammavsn} can be estimated as the average energy of the IceCube and DeepCore event distributions,
$\langle E \rangle$, as
\be
\langle E \rangle \equiv 
\frac{\displaystyle \int \frac{dN}{dE}E dE}{\displaystyle
\int\frac{dN}{dE} dE}\ ,
\label{eq:average}
\ee
where $dN/dE$ is the event number distribution. 
This leads to $\langle E \rangle\simeq 4$ TeV ($40$ GeV) for IceCube (DeepCore), which are in the right ballpark 
although somewhat different from the values of $E_0$ giving the best fit to the data shown in fig.~\ref{fig:Bounds-vs-n}. 
Nevertheless, we find these to be in reasonable agreement, given our naive estimation of $E_0$ as the mean energy for each experiment.

\section{Summary and Conclusions}
\label{sec:conclusions}

\begin{table*}[ht!]
\small
\centering
\tabcolsep=0.11cm
\renewcommand{\arraystretch}{1.3}
\begin{tabular}{| c | c | ccccc |}
\hline
\multirow{8}{*}{\begin{sideways} Normal Ordering \phantom{aaaaaaaa}  \end{sideways}} & & $n=-2$ & $n=-1$ & $n=0$ & $n=1$ & $n=2$\cr
\hline
\hline
& \textbf{IceCube (this work)}  & & & & &\cr
\cline{2-7}
& Atmospheric ($\gamma_{31}=\gamma_{32}$) &$2.8\cdot10^{-18}$ & $4.2\cdot10^{-21}$  & $\mathbf{4.0\cdot10^{-24}}$  &  $\mathbf{1.0\cdot10^{-27}}$ &  $\mathbf{1.0\cdot10^{-31}}$\cr
& Solar I ($\gamma_{31}=\gamma_{21}$)    & $6.8\cdot10^{-19}$ & $1.2\cdot10^{-21}$  & $\mathbf{1.3\cdot10^{-24}}$ &  $\mathbf{3.5\cdot10^{-28}}$&   $\mathbf{1.9\cdot10^{-32}}$ \cr 
& Solar II ($\gamma_{32}=\gamma_{21}$)    &  $5.2\cdot10^{-19}$  & $9.2\cdot10^{-22}$   & $9.7\cdot10^{-25}$ & $2.4\cdot10^{-28}$ & $9.0\cdot10^{-33}$  \cr
\cline{2-7}
& \textbf{DeepCore (this work)} & & & & &\cr
\cline{2-7}
& Atmospheric ($\gamma_{31}=\gamma_{32}$)  &  $\mathbf{4.3\cdot10^{-20}}$ & $2.0\cdot10^{-21}$ & $8.2\cdot10^{-23}$ & $3.0\cdot10^{-24}$ & $1.1\cdot10^{-25}$\cr
& Solar I  ($\gamma_{31}=\gamma_{21}$)    & $ \mathbf{1.2\cdot 10^{-20}}$ & $  5.4\cdot10^{-22}$ & $ 2.1\cdot 10^{-23}$ &  $6.6\cdot10^{-25}$&  $2.0\cdot10^{-26}$  \cr 
& Solar II ($\gamma_{32}=\gamma_{21}$)    & $7.5\cdot 10^{-21}$ & $3.5\cdot 10^{-22}$  & $ 1.4\cdot 10^{-23}$ & $4.2\cdot 10^{-25}$  &  $ 1.1\cdot 10^{-26}$ \\ 
\hline
\hline
\multirow{7}{*}{\begin{sideways} Inverted Ordering \phantom{aaaaa}  \end{sideways}}  &\textbf{IceCube (this work)}  & & & & & \cr
\cline{2-7}
& Atmospheric ($\gamma_{31}=\gamma_{32}$) & $6.8\cdot10^{-19}$ & $1.2\cdot10^{-21}$  & $\mathbf{1.3\cdot10^{-24}}$ &  $\mathbf{3.5\cdot10^{-28}}$& $\mathbf{1.9\cdot10^{-32}}$ \cr
& Solar I ($\gamma_{31}=\gamma_{21}$) &  $5.2\cdot10^{-19}$  & $9.2\cdot10^{-22}$  & $9.8\cdot10^{-25}$ & $2.4\cdot10^{-28}$ & $9.0\cdot10^{-33}$  \cr  
& Solar II ($\gamma_{32}=\gamma_{21}$) &$2.8\cdot10^{-18}$ & $4.2\cdot10^{-21}$ & $\mathbf{4.1\cdot10^{-24}}$  &  $\mathbf{1.0\cdot10^{-27}}$ &  $\mathbf{1.0\cdot10^{-31}}$\cr   
\cline{2-7}
&\textbf{DeepCore (this work)} & & & & &\cr
\cline{2-7}
& Atmospheric ($\gamma_{31}=\gamma_{32}$) &  $\mathbf{1.4\cdot 10^{-20}}$  & $  5.8\cdot10^{-22}$ & $ 2.2\cdot 10^{-23}$ &  $7.5\cdot10^{-25}$&  $2.3\cdot10^{-26}$  \cr 
& Solar I ($\gamma_{31}=\gamma_{21}$) & $8.3\cdot 10^{-21}$  & $3.6\cdot 10^{-22}$ & $ 1.4\cdot 10^{-23}$ & $4.7\cdot 10^{-25}$  &  $ 1.3\cdot 10^{-26}$ \cr 
& Solar II ($\gamma_{32}=\gamma_{21}$) &  $\mathbf{5.0\cdot10^{-20}}$  & $2.3\cdot10^{-21}$ & $9.4\cdot10^{-23}$ & $3.3\cdot10^{-24}$ & $1.2\cdot10^{-25}$\\  
\hline
\hline 
& \textbf{Previous Bounds} & & & & & \\
\hline
& SK (two families)~\cite{Lisi:2000zt}  & &$2.4\cdot 10^{-21}$ & $4.2\cdot10^{-23}$ & &  $1.1\cdot10^{-27}$\\ 
& MINOS ($\gamma_{31},\gamma_{32}$)~\cite{deOliveira:2013dia} & & $\mathbf{2.5 \cdot 10^{-22}}$ &$1.1\cdot 10^{-22}$ & $2\cdot10^{-24}$ & \\ 
& KamLAND ($\gamma_{21}$)~\cite{Gomes:2016ixi}    &  & $\mathbf{3.7\cdot 10^{-24}}$ & $6.8\cdot 10^{-22}$ & $1.5\cdot10^{-19}$&  \\  
\hline
\end{tabular}
\caption{DeepCore/IceCube bounds on $\gamma_{ij}^0$ in GeV ($\gamma_{ij}=\gamma_{ij}^0(E/\text{GeV})^n$), at the $95\%$ CL (1 degree of freedom), for both normal and inverted ordering as indicated. Previous constraints are also provided for comparison, and the dominant limit in each case is highlighted 
in bold face (notice that we considered the most conservative bound from the two solar limits).}
\label{tab:summary}
\end{table*}

In this work, we have derived strong limits on non-standard neutrino decoherence
parameters in both the solar and atmospheric sectors from 
the analysis of IceCube and DeepCore atmospheric neutrino data. 
Our analysis includes matter effects in a consistent manner within a three-family 
oscillation framework, unlike most past literature on this topic. In sec.~\ref{sec:formalism} we have developed a general 
formalism, dividing the matter profile into layers of constant density, which permits to study decoherence effects in neutrino oscillations 
affected by matter effects in a non-adiabatic regime. 
Our analysis shows that the matter effects are extremely relevant for atmospheric 
neutrino oscillations and their importance in order to correctly interpret the two-family limits obtained previously in the literature, as outlined in sec.~\ref{sec:plots-probabilities}.

We have found that the sensitivity to decoherence effects depends strongly on the
neutrino mass ordering and on whether the sensitivity 
is dominated by the neutrino or antineutrino event rates. For neutrinos, the decoherence effects at high energies 
are mainly driven by $\gamma_{21}$ ($\gamma_{31}$) for normal (inverted) ordering, while in the antineutrino case they are essentially controlled by 
$\gamma_{32}$ ($\gamma_{21}$) for normal (inverted) ordering. This means that, considering a three-family framework including matter effects is
essential when decoherence effects in atmospheric neutrino oscillations are studied. Our results are summarized in tab.~\ref{tab:summary}, together with the most relevant bounds present in the literature. Table~\ref{tab:summary} provides the 95\% CL bounds extracted from our analysis of DeepCore and IceCube atmospheric neutrino data, for both normal and inverted ordering,
and for the three limiting cases considered in this work: (A) atmospheric limit ($\gamma_{21}=0$), (B) solar limit I ($\gamma_{32}=0$) and (C) 
solar limit II ($\gamma_{31}=0$). In~\ref{sec:d5results} we show that the bounds derived in these limits correspond
to the most conservative results that can be extracted in the general case. 

In this work, we considered a general dependence of the decoherence parameters with the energy, as  
$\gamma_{ij}=\gamma_{ij}^0\left(E/\text{GeV}\right)^n$ with $n=\pm2,0,\pm2$. Our results improve over previous bounds for most of the cases studied, with the exception of the $n=-1$ case. For $n=-1$, KamLAND gives the dominant
bound on $\gamma_{21}$ while MINOS gives the strongest constraints on $\gamma_{31}$ and $\gamma_{32}$\footnote{Reactor experiments as Double Chooz, Daya Bay 
or RENO are also expected to give a competitive bound in the atmospheric sector, as it was shown in~\cite{An:2016pvi} for Daya Bay in the standard decoherence case.} (indeed, both KamLAND and MINOS are also expected to 
give the strongest bound for $n=-2$, although to the best of our knowledge no analysis has been performed for this case yet). 
We have found that DeepCore considerably improves the present bounds in the solar sector ($\gamma_{21}$) for $n=0,1,2$ and gives a constraint in the atmospheric sector 
comparable to the SK limit, although a factor $2$ weaker, in the $n=0$ case. Our results show that, for $n=0$ (which is the case most commonly considered in the literature), 
IceCube improves the bound on $\gamma_{31}$ and $\gamma_{32}$ in (more than) one order of magnitude with respect to the SK constraint,
obtained in a simplified two-family approximation, and by more than one order (almost two orders) of 
magnitude for NO (IO) with respect to the KamLAND constraint on $\gamma_{21}$. 
In particular, we find that the reference value for $\gamma_{23}$ considered in ref.~\cite{Coelho:2017zes} to explain the small tension previously 
reported between NOvA and SK data is indeed already excluded by IceCube 
data. Regarding the cases with $n = 1,2$, we find that the sensitivity of IceCube is particularly strong. 
For instance, IceCube improves the bound from KamLAND on $\gamma_{21}$
by almost 9 (8) orders of magnitude for $n=1$ and NO (IO), while for $n=2$ the bound on $\gamma_{31}$ and $\gamma_{32}$ is 
improved in 4 (5) orders of magnitude with respect to the SK limit for NO (IO).

\begin{acknowledgements}

We thank M.~C.~Gonz\'alez-Garc\'ia, M.~Maltoni and J.~Salvado for
useful discussions. JLP, IMS and HN thank the hospitality of the
Fermilab Theoretical Physics Department where this work was
initiated. HN also thanks the hospitality of the CERN Theoretical
Physics Department where the final part of this work was done. HN was
supported by the Brazilian funding agency, CNPq (Conselho Nacional de
Desenvolvimento Cient\i{\i}fico e Tecnol\'ogico), and by Fermilab
Neutrino Physics Center. IMS acknowledges support from the Spanish
grant FPA2015-65929-P (MINECO/FEDER, UE) and the Spanish Research
Agency (``Agencia Estatal de Investigacion'') grants IFT ``Centro de
Excelencia Severo Ochoa'' SEV2012-0249 and SEV-2016-0597. This work
was partially supported by the European projects
H2020-MSCA-ITN-2015-674896-ELUSIVES and
690575-InvisiblesPlus-H2020-MSCA-RISE-2015.This manuscript has been
authored by Fermi Research Alliance, LLC under Contract
No. DE-AC02-07CH11359 with the U.S. Department of Energy, Office of
Science, Office of High Energy Physics. The publisher, by accepting
the article for publication, acknowledges that the United States
Government retains a non-exclusive, paid-up, irrevocable, world-wide
license to publish or reproduce the published form of this manuscript,
or allow others to do so, for United States Government purposes.
\end{acknowledgements}

\appendix

\section{Computation of oscillation probabilities in three layers }
\label{app}

The simulation of atmospheric neutrino experiments is computationally demanding in the standard three-family scenario, and even more if decoherence effects are included in the analysis. Therefore, due to the cost of implementing a large number of layers for the PREM profile
density, in this work we consider a simplified three-layer model for 
the Earth matter density profile assuming a core and Earth radii of 3321 km and 6371 km, respectively. The values of the matter densities of the inner layer
(core) and the outer layer (mantle) are taken to be around 
$\rho = 12$ g/cm$^3$ and $4.6$ g/cm$^3$, respectively. However, their values 
are slightly adjusted depending on 
the distance traveled by the neutrinos to match as close as possible 
the profile of the PREM model~\cite{Dziewonski:1981xy}. Note that, in our simulations, we have not considered the atmosphere 
as an additional layer. This is a good approximation for neutrinos going upwards in the detector ($\cos\theta_z < 0$), but is 
not a valid approximation in the region $\cos\theta_z >0$. This has only an impact on the IceCube results for extremely large values 
of the decoherence parameters, which are already ruled out either by other experiments or by DeepCore.

\begin{figure*}[h!]
  \centering
\includegraphics[width=1.98\columnwidth]{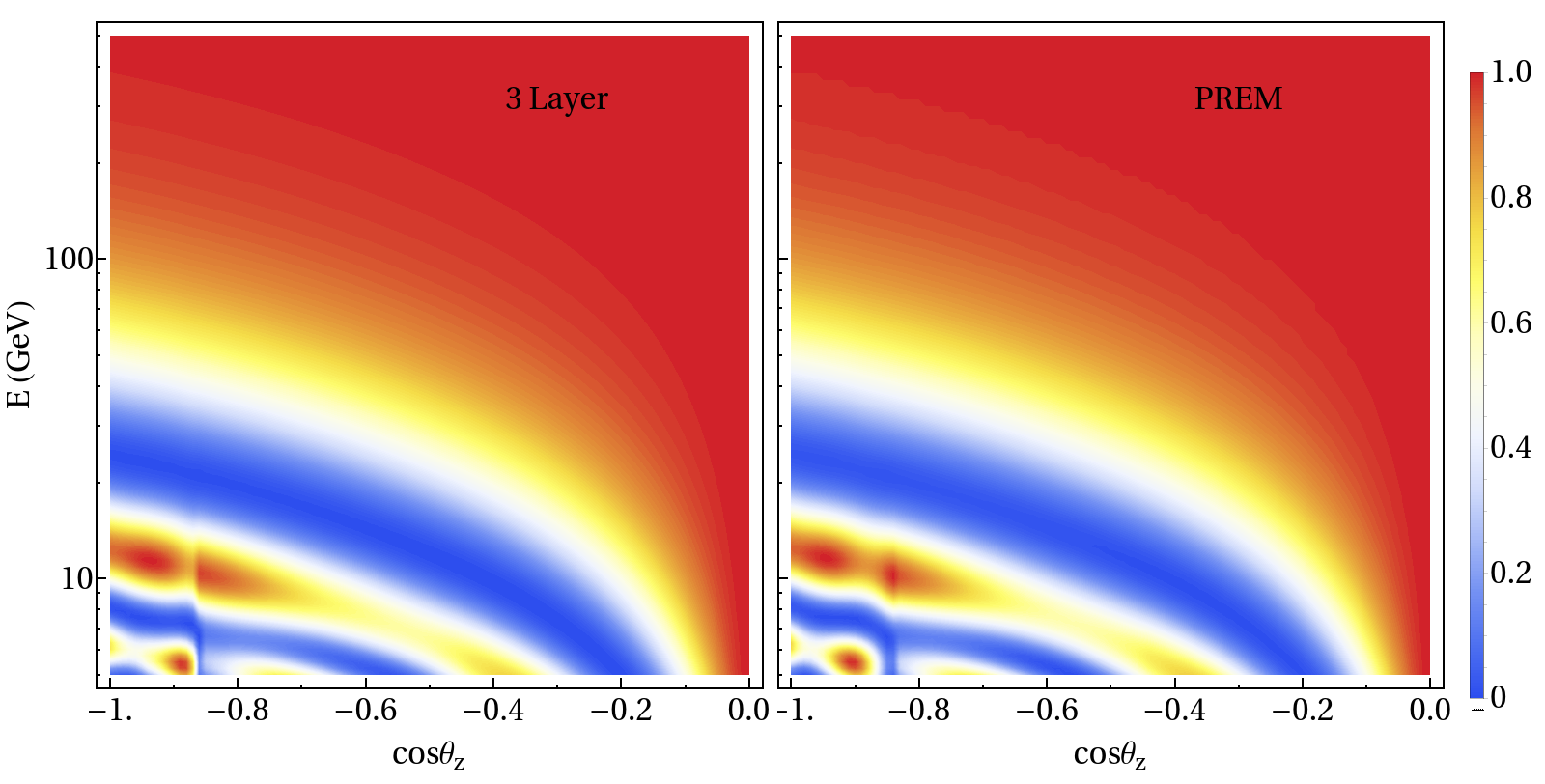}
  \caption{Oscillograms for $P_{\mu\mu}$ without decoherence, considering our three-layer approximation (left panel) and the PREM model 
  (right panel) for the Earth matter density profile.}
  \label{fig:earthdens}
\end{figure*}

In fig.~\ref{fig:earthdens} we compare the results obtained for the oscillation probability for our modified three-layer approximation (left panel) 
against the exact numerical results using the full Preliminary Reference Earth Model (PREM) profile~\cite{Dziewonski:1981xy} (right panel), which divides the Earth 
into eleven layers where the matter density in each layer is given by a polynomial function of the distance traveled. In this figure, the results are shown for the standard three-family 
scenario with no decoherence, in order to illustrate the accuracy of our three-layer approximation. The results are shown as a 
neutrino oscillogram, which represents the oscillation probability in the $P_{\mu\mu}$ channel in 
terms of energy and the zenith angle $\theta_z$ of the incoming neutrino. In this figure, 
a normal mass ordering was assumed, together with the following input values for the oscillation parameters~\cite{nufit,Esteban:2016qun}: $\Delta m^2_{21} = 7.4 \cdot 10^{-5}$ eV$^2$, 
$\Delta m^2_{31} = 2.515 \cdot 10^{-3}$ eV$^2$, 
$\theta_{12} = 33.62^\circ$, 
$\theta_{13} = 8.54^\circ$, 
$\sin^2\theta_{23} = 0.51$, 
and $\delta = 234^\circ$. 

As can be seen from the comparison between the two panels, some small differences take place but only in a restricted range of values of energy and zenith angle. Therefore, we conclude that the agreement between the probabilities obtained using the exact PREM model (right) and our approximate three-layer model (left) is sufficiently
good for the purposes of this work. We 
have also checked that, using our simplified three-layer model applied to the standard case without decoherence,
we are able to reproduce up to a very good approximation the DeepCore oscillation fit for the atmospheric 
parameters $\theta_{23}$ and $\Delta m^2_{32}$~\cite{Aartsen:2014yll}.

\section{Five-dimensional analysis}
\label{sec:d5results}

The $\gamma_{ij}$ are not completely independent parameters, see eq.~(\ref{eq:gamma_dm}). 
In order to simplify the analysis, in this work we have focused on three different representative cases: (A) Atmospheric limit, $\gamma_{21}=0$ 
($\gamma_{32} = \gamma_{31}$); (B) Solar limit I, $\gamma_{32}=0$ ($\gamma_{21} = \gamma_{31}$); and (C) Solar limit II, 
$\gamma_{31}=0$ ($\gamma_{21} = \gamma_{32}$). Considering these one-$\gamma_{ij}$-dominated cases is expected to be a very good approximation in view 
of eqs.~\eqref{eq:NOnu}-\eqref{eq:IOnubar}. Nevertheless, in this appendix we will show that the results obtained in these simplified
scenarios also apply to the more general case in which the three $\gamma_{ij}$ are different from zero.

\begin{figure*}[ht!]
  \centering
  \includegraphics[width=1.9\columnwidth]{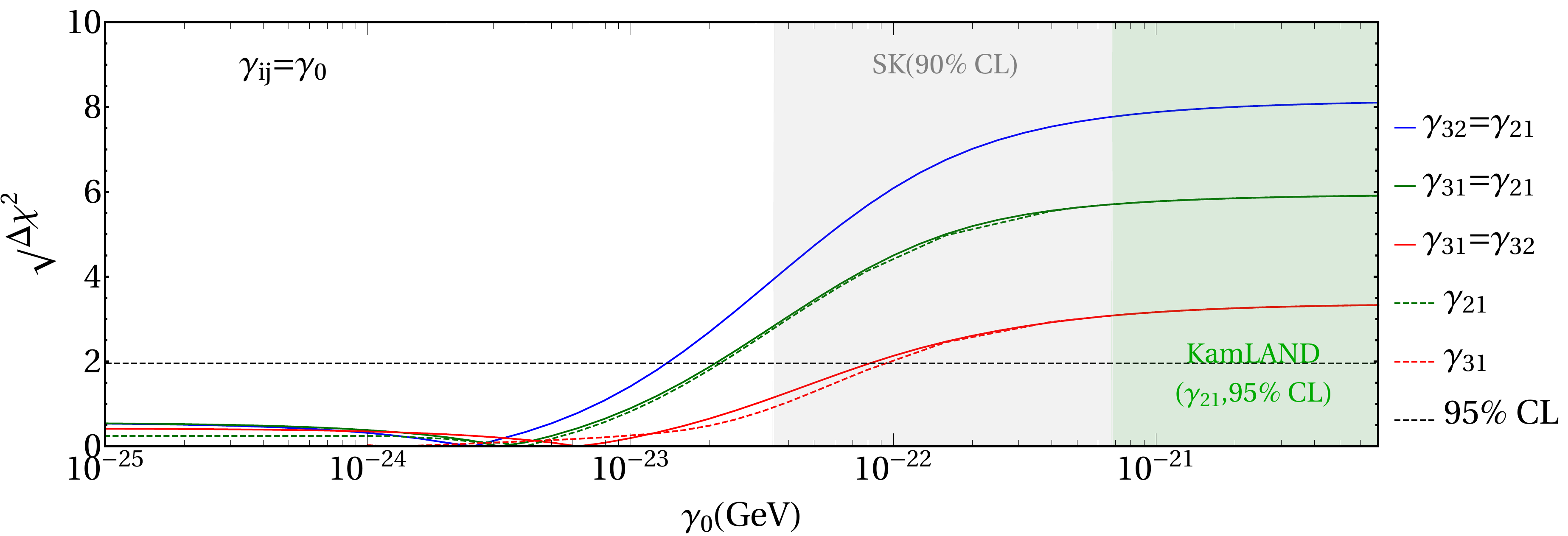}
  \caption{ $\sqrt{\Delta\chi^2}$ obtained from the five-dimensional DeepCore analysis as a function of 
$\gamma_{21}$ (dashed green curve) and $\gamma_{31}$ (dashed red curve), marginalizing over the rest of 
the free parameters, for $n=0$ and NO. The $\sqrt{\Delta\chi^2}$ for the Atmospheric (solid red curve), 
Solar I (solid green curve) and Solar II (solid blue curve) limits is also shown.}
  \label{fig:5Param}
\end{figure*}

Let us assume that just one $D_{m}$ matrix contributes to the decoherence term of the evolution equations given by 
eq.~(\ref{eq:Dm}). In such a case, one of the $\gamma_{ij}$ parameters is a function of the other
two $\gamma_{ij}$. Without loss of generality, if we choose $\gamma_{21}$ and $\gamma_{31}$ as our free parameters, 
$\gamma_{32}$ is then given by  
\be \gamma_{32} = \left(\sqrt{\gamma_{21}}\pm
\sqrt{\gamma_{31}}\right)^2.
\ee
In order to understand how general are the results presented in sec.~\ref{sec:results}, we have performed a five-dimensional analysis varying 
$\gamma_{21},\,\gamma_{31},\,\theta_{23}$ and $\Delta m^2_{32}$ in the fit, and imposing the constraint given by the equation above.
In fig.~\ref{fig:5Param} we show the $\sqrt{\Delta\chi^2}$ obtained from the five-dimensional DeepCore analysis as a function of 
$\gamma_{21}$ (dashed green curve) and $\gamma_{31}$ (dashed red curve), marginalizing over the rest of 
the free parameters, for the $n=0$ case (the same conclusions apply to the other cases studied in this work). For the sake of comparison,
the $\sqrt{\Delta\chi^2}$ associated to the atmospheric (solid red curve), solar I (solid green curve) and solar II (solid blue curve) limits  is 
also included in the same figure. NO was assumed but the results can be easily extrapolated to the IO case using the
mapping given in~eq.~(\ref{eq:map}).

Figure~\ref{fig:5Param} shows that the five-dimensional $\sqrt{\Delta\chi^2}$ 
distribution projected into $\gamma_{31}$ coincides with the Atmospheric limit one, while 
when it is projected into $\gamma_{21}$ resembles the most conservative of the two 
solar limits. This is due to the marginalization over the parameters which are not shown. For instance, in the case of $\gamma_{21}$ the marginalization selects, between the two solar limits, 
the most conservative result. We conclude therefore 
that our analysis distinguishing the three limits (A), (B) and (C), provides the most conservative bounds
that can be applied to the general case in which the three $\gamma_{ij}$ are different from zero. 
 
\providecommand{\href}[2]{#2}
\begingroup\raggedright
\endgroup

\end{document}